\documentclass[a4paper,fleqn,usenatbib]{mnras}

\usepackage{graphicx}	
\usepackage{amsmath}	
\usepackage{amssymb}	
\usepackage{longtable}
\usepackage{hyperref}
\usepackage[hyphenbreaks]{breakurl}

\def\teff{$T\rm_{eff}$}

\def\logg{$\log\,g$}
\def\loggf{$\log\,gf$\ }

\title[Departures from LTE in unevolved stars]
{
Study of the departures from LTE in the unevolved stars infra-red spectra.
\thanks{Based on observations made with GIANO at the Italian Telescopio Nazionale
Galileo (TNG) operated on the island of La Palma by the Fundaci\'on Galileo
Galilei of the INAF (Istituto Nazionale di Astrofisica) at the Spanish
Observatorio del Roque de los Muchachos of the Instituto de Astrofisica de
Canarias}
}

\author[S.A. Korotin et al.]{
S.A. Korotin,$^{1}$\thanks{E-mail: serkor1@mail.ru},
S.M. Andrievsky,$^{2,3}$ 
E. Caffau,$^{3}$
P. Bonifacio$^{3}$
and E. Oliva$^{4}$
\\
$^{1}$Crimean Astrophysical Observatory, Nauchny 298409, Republic of Crimea\\
$^{2}$Astronomical Observatory, Odessa National University, Shevchenko Park, 65014, Odessa, Ukraine\\
$^{3}$GEPI, Observatoire de Paris, PSL Research University, CNRS, Place Jules Janssen, 92195 Meudon, France\\
$^{4}$INAF-Osservatorio Astrofisico di Arcetri, Largo E. Fermi 5, 50125 Firenze, Italy\\
}

\date{Accepted XXX. Received YYY; in original form ZZZ}

\pubyear{2020}

\begin{document}
\label{firstpage}
\pagerange{\pageref{firstpage}--\pageref{lastpage}}
\maketitle

\begin{abstract}
We present a study of departures from Local Thermodynamic Equilibrium (LTE) in 
the formation of infra-red lines of Na\,{\sc i}, Mg\,{\sc i}, Al\,{\sc i}, S\,{\sc i}, 
K\,{\sc i} and Sr\,{\sc ii} in unevolved stars of spectral types F,G,K and metallicities 
around the solar metallicity. The purpose of this investigation
is to identify lines of these species  that can be safely treated with the
LTE approximation in the infra-red  spectra of these types of stars.
We employ a set of 40 stars observed with the GIANO spectrograph
at the 3.5\,m Telescopio Nazionale Galileo (TNG) and previously
investigated by Caffau et al.
We were able to identify many lines that can be treated in LTE for all
the above-mentioned species, except for Sr\,{\sc ii}. The latter species
can only be studied using three lines in the J-band, but all three of them
display significant departures from LTE. 
With our small-size, but high-quality sample we can determine
robustly the trends of the abundance ratios with metallicity, 
confirming the trends apparent from a sample that
is larger by several orders of magnitude, but of lower quality in terms
of resolution and S/N ratio.

\end{abstract}





\section{Introduction}

With the advent of Gaia and  especially the Gaia data release 2 \citep[DR2,][]{GaiaDR2,Arenou}
our vision on the Milky Way has changed: our Galaxy revealed to be bigger, the stars on average
are more distant from the Sun than what we thought. 
For example, the classical vision we had of the Galactic Halo formed by old, metal-poor, ``in situ'' stars has changed:
this extended Galacic component is likely populated by stars accreted in a major event 
\citep[see e.g.][]{belokurov18,haywood18,helmi18,dimatteo19}.
However, the kinematics alone is not able to distinguish between
the accreted and the ``in situ'', population \citep{dimatteo19}. 
Detailed, high-precision chemical inventory of a stellar population can do the job,
especially in the high metallicity regime \citep[see e.g.][]{NS10,NS11}.
The Gaia RVS will provide metallicites and a limited set of chemical abundances in the future
Gaia releases, but the sample will be magnitude limited. Past \citep[e.g.][Gaia-UVES Survey]{gesgg}, 
on-going \citep[e.g.][APOGEE]{apogee08} and future spectroscopic surveys \citep[e.g.][WEAVE and 4MOST]{WEAVE,4MOST} provided/are providing/will provide the detailed chemical informations missing from Gaia.

In recent years, a deep interest in the near infrared (NIR) ranges was evident.
Several spectrographs have been built or are under construction to observe the NIR ranges.
A non-exhaustive list includes:
\begin{itemize}
\item
CRIRES \citep{Kaeufl04} operated at the UT1, Antu telescope of the VLT, from 2006 to 2014,
observing four intervals of the order of 5\,nm wide, contained in the range $955\leq 5248$\,nm. 
The upgrade CRIRES will soon be available at VLT, with a higher capability in wavelength range coverage.
\item
APOGEE \citep{apogee08} is a multi-object spectrograph, covering a wavelength range in the H-band (1510--1700\,nm), observing at intermediate resolution (R$\approx 22500$).
\item
Multi Object Optical and Near-infrared Spectrograph \citep[MOONS,][]{moons14},
that will provide high resolution (R = 19\,700)  H-band spectra 
for 1000 targets over a field of view of $20'$ diameter at each pointing,
as well as lower resolution I-band and J-band spectra.
\item
GIANO \citep{giano14} is a near-infrared (NIR) cross-dispersed echelle spectrograph, operating at Telescopio Nazionale Galileo (TNG).
It covers the wavelength range 950--2450\,nm and operate at high-resolving power (R$\approx 50\,000$).
\item
SPiRou\footnote{http://spirou.irap.omp.eu/}
is a near-infrared (NIR) cross-dispersed echelle spectrograph, operating at Canada-France-Hawaii Telescope.
\item
CARMENES \citep{CARMENES} is a cross-dispersed echelle spectrograph with two arms:  visible and  NIR.
\item
IRTF  (InfraRed Telescope Facility) 3-meter telescope on Maunakea equipped with cross-dispersed echelle spectrograph  iShell (1.08--5.3 micron, R = 80000, \citealt{Rayner16}).
\end{itemize}

APOGEE is providing the community with a large amount of spectra 
\citep[its release 16 provides more than 280\,000 stellar spectra][]{apogeedr16} and detailed chemical 
investigation \citep[see e.g.][]{weinberg19} for a signigficant sample of elements.
However, the devil is in the details, to compare chemical abundances of stars spanning large ranges in
effective temperature, derived in different spectral ranges, like visible and IR,
to provide conclusions on the chemical evolution of the Galaxy, 
one should try to be as careful as possible. The departure from Local Thermodynamycal Equilibrium (NLTE) 
can play a role in affecting in different ways stars with different stellar parameters.

This paper is methodological in nature, and investigates how departures from
LTE  affect the 
individual lines of some chemical elements (Na, Mg, Al, S, K and Sr) in the infrared wavelength domain.
For this purpose, we use the high-quality spectra of 40 dwarfs obtained with 
the GIANO spectrograph,
already analysed by \citet{Caffau19a} to derive stellar parameters and chemical compositions by using 
standard LTE approximation. Nevertheless, \citet{Caffau19a}
have also made the first attempt to evaluate the role of  NLTE effects on 
the derived abundances, at least for a few elements (Na, Mg, Al, K and Ca). We decided to revisit 
the abundances of those program stars and in this study we investigate the NLTE effects on the individual 
lines. We hope that our NLTE results will help other 
specialists to have the chance to provide careful abundance in their analysis based on IR spectra 
while using LTE approximation instead of NLTE analysis. We provide informations on 
which lines can be suitable for classic LTE analysis, producing quite reliable abundance results.
To check the limits of applicability of our conclusions about the possibility 
of using the LTE approximation to derive the elemental abundances, we carried 
out an additional study of the problem by NLTE analysis for the more expand 
field of atmospheric parameters. Qualitative results are presented in the 
Discussion section for each of the elements studied.

At our knowledge, no systematic investigation on NLTE has been done in the GIANO IR spectral domain.
We here present results of our investigations. Among our sample of stars observed with GIANO  there are 
12 stars for which optical  archive spectra are available (from various spectrographs, 
UVES, ESPaDOnS, Narval and Sophie). All these spectra were also used for comparison of the abundances 
derived from the IR and optical spectra.


\section{Method of NLTE analysis and individual ions}

We first checked our NLTE atomic models to verify how adequately they reproduce the 
profiles of the IR lines of interest in the solar spectrum. The primary aim was 
to check the accuracy of oscillator strengths and damping parameters which are needed for calculations. 
We adopted the solar abundances derived in previous investigations from optical lines (see Table\,\ref{sun}).
We then investigated each line analysed in this work in the solar specrum by comparing solar observations
to the NLTE synthesis with the solar abundance 
derived from optical spectra and adopted here.
The best choice would have been to use for this aim the solar 
spectrum observed with GIANO. Unfortunately, we had not such a spectrum in our 
disposal, therefore, we used solar atlases of \citet{Kurucz84} (region $<$ 
12500 \AA), \citet{Reiners16} (region 12500--22500 \AA), ACE-FTS \citet{Hase10} 
(region $>$ 22600 \AA), NSO/Kitt Peak FTS \citet{Livingston91} (region 12500--22500 \AA).
We then adjusted the atomic data of each IR line here studied in a way to have a good agreement between 
the NLTE synthesis and the observed solar spectrum.

\begin{table}
\caption{Solar abundances adopted here.}
\label{sun}
\centering
\begin{tabular}{lr}
\hline
\hline
Element & A(X) \\
\hline
Na      & 6.25 \\
Mg      & 7.54 \\
Al      & 6.43 \\
S       & 7.16 \\
K       & 5.11 \\
Sr      & 2.92 \\
Fe      & 7.52 \\
\hline                                                                                            
\end{tabular}                                                                               
\end{table}

In order to find atomic level populations for the following ions: \ion{Na}{I}, 
\ion{Mg}{I}, \ion{Al}{I}, \ion{S}{I}, \ion{K}{I} and \ion{Sr}{II}, we employed 
the code MULTI \citep{Carlsson86}. For our aim this program was modified and 
adapted by \citet{Korotin99}. MULTI enables one to calculate a single line NLTE profile. 
Nevertheless, the lines of interest are often blended in the real stellar spectra. In order to take 
the blending into account, we first calculate with MULTI the departure coefficients 
for those levels that form the line of interest, and then we include these 
coefficients in the LTE synthetic spectrum code SYNTHV \citep{Tsymbal96}. This
allows one to calculate the source function and opacity for each studied line. 
Simultaneously, the blending lines are calculated in LTE with the
help of line list and corresponding atomic data from VALD data base 
\citep{Ryabchikova15} in the wavelength range of the line under study.

\begin{figure}
\includegraphics[width=0.98\columnwidth,clip=true]{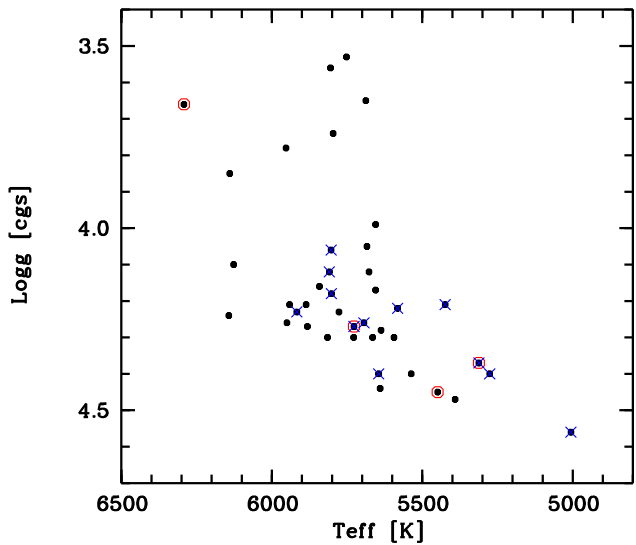}
\caption{The stellar parameters of the complete sample (filled black circles).
The four stars for which the LTE--NLTE comparison has been made have a red circle around the black symbol.
The twelve stars for which we performed an optical inversigation are highlighted with a blue cross.}
\label{fig:hr}
\end{figure}

For all our computations we have used 1-D LTE atmosphere models computed 
with ATLAS9 code by \citet{Kurucz93,Kurucz05}. For our program 
stars we used atmosphere parameters derived by \citet{Caffau19a} with 
microturbulence velocity of 1\,km~s$^{-1}$.  The atmospheric parameters \teff, \logg, [Fe/H] 
are presented in Table \ref{ab}, where we adopted for the Sun ${\rm A}({\rm Fe})=7.52$ \citep{caffau11}.
The twelve stars, for which the NLTE abundances were also
determined with the help of optical spectra, are marked in Table\,\ref{ab} with asterisks.
The lines and atomic data used in the optical investigation of Na, Mg, Al, S, K 
and Sr for these twelve stars were described in each introductive papers 
devoted to the NLTE atomic model for each ion.
In Fig.\,\ref{fig:hr} the sample of stars is presented in the  \teff, \logg\ plane; the twelve stars above mentioned
are highlighted with blue crosses.

\begin{figure}
\includegraphics[width=0.98\columnwidth,clip=true]{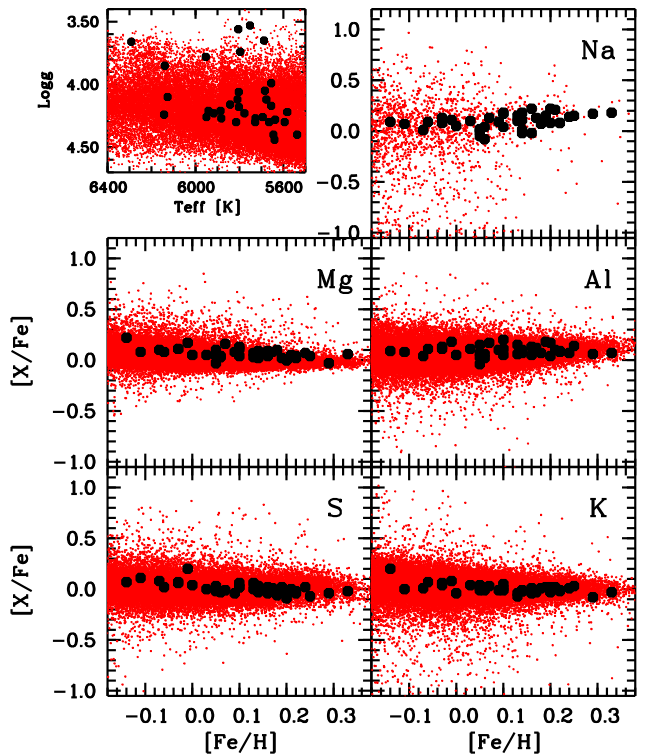}
\caption{The parametrs and abundances of the stars here analyzed (black filled circles)
compared to the APOGEE\,DR16 data.}
\label{apogee_comp}
\end{figure}

In the following sections we shall compare our abundances both
with those of \citet{Caffau19a} for the same set of stars
and with the abundances of a set of 41\,552 stars with stellar
parameters similar to our sample extracted from the APOGEE catalogue, 
SDSS data release 16 \citep{apogeedr16}.
The summary plots of this latter comparison are shown in Fig.\,\ref{apogee_comp}.

\begin{table*}
\caption{Atmosphere parameters of program stars and our NLTE abundances of the elements analyzed.}
\label{ab}
\centering
\begin{tabular}{lccrrrrrrr}
\hline
\hline
  Star  & T$_{\rm eff}$, K&log g&[Fe/H]&[Na/Fe]       &[Mg/Fe]        &[Al/Fe]        &[S/Fe]        &[K/Fe]        &[Sr/Fe]    \\
\hline                                                                                        
HD 20670& 5688& 3.65& 0.10 & 0.17$\pm$0.10&  0.11$\pm$0.05&  0.11$\pm$0.07& 0.02$\pm$0.06& 0.05$\pm$0.12& --0.03$\pm$0.03\\
HD 24040*& 5809& 4.12& 0.09 & 0.09$\pm$0.12&  0.08$\pm$0.05&  0.10$\pm$0.10&--0.04$\pm$0.08&--0.01$\pm$0.12& --0.05$\pm$0.02\\
HD 28005*& 5802& 4.18& 0.21 & 0.21$\pm$0.10&  0.05$\pm$0.07&  0.10$\pm$0.10& 0.00$\pm$0.07& 0.02$\pm$0.12& --0.08$\pm$0.03\\
HD 32673& 5752& 3.53& 0.07 & 0.13$\pm$0.10&  0.16$\pm$0.05&  0.17$\pm$0.07&--0.01$\pm$0.07& 0.03$\pm$0.10&  0.01$\pm$0.04\\
HD 34445*& 5803& 4.06&--0.03 & 0.13$\pm$0.10&  0.11$\pm$0.05&  0.11$\pm$0.10& 0.07$\pm$0.07& 0.04$\pm$0.10&  0.06$\pm$0.03\\
HD 34575*& 5582& 4.22& 0.18 & 0.07$\pm$0.10&  0.10$\pm$0.06&  0.07$\pm$0.08&--0.06$\pm$0.08& 0.01$\pm$0.12& --0.05$\pm$0.05\\
HD 44420& 5777& 4.23& 0.19 & 0.15$\pm$0.13&  0.02$\pm$0.05&  0.18$\pm$0.12& 0.02$\pm$0.10& 0.02$\pm$0.12&  0.00$\pm$0.03\\
HD 56303& 5941& 4.21& 0.05 & 0.02$\pm$0.10&  0.07$\pm$0.06&  0.07$\pm$0.08&--0.02$\pm$0.08&--0.01$\pm$0.10&  0.08$\pm$0.04\\
HD 67346& 5953& 3.78& 0.14 & 0.11$\pm$0.15&  0.07$\pm$0.06&  0.07$\pm$0.10& 0.03$\pm$0.10&--0.02$\pm$0.10&  0.07$\pm$0.04\\
HD 69056& 5637& 4.28& 0.05 & 0.01$\pm$0.10&  0.10$\pm$0.05&  0.15$\pm$0.10&--0.01$\pm$0.10& 0.04$\pm$0.10&  0.00$\pm$0.05\\
HD 69809& 5842& 4.16& 0.17 & 0.13$\pm$0.12&  0.08$\pm$0.06&  0.12$\pm$0.08& 0.01$\pm$0.08&--0.03$\pm$0.10& --0.06$\pm$0.03\\
HD 69960& 5655& 3.99& 0.22 & 0.08$\pm$0.10&  0.02$\pm$0.06&  0.08$\pm$0.10&--0.02$\pm$0.08&--0.01$\pm$0.10& --0.08$\pm$0.05\\
HD 73226& 5886& 4.21& 0.06 & 0.04$\pm$0.12&  0.03$\pm$0.06&  0.06$\pm$0.10&--0.02$\pm$0.07& 0.01$\pm$0.08&  0.03$\pm$0.01\\
HD 73933& 6143& 4.24& 0.06 &-0.08$\pm$0.12&  0.04$\pm$0.05&  0.02$\pm$0.10&--0.03$\pm$0.08&--0.01$\pm$0.10&  0.14$\pm$0.06\\
HD 76909& 5655& 4.17& 0.24 & 0.14$\pm$0.15&  0.07$\pm$0.07&  0.07$\pm$0.08& 0.02$\pm$0.10& 0.01$\pm$0.10&  0.00$\pm$0.07\\
HD 77519& 6140& 3.85& 0.14 & 0.01$\pm$0.12&  0.08$\pm$0.06&  0.09$\pm$0.15& 0.03$\pm$0.12&--0.01$\pm$0.10&  0.14$\pm$0.05\\
HD 82943*& 5917& 4.23& 0.13 & 0.10$\pm$0.12&  0.03$\pm$0.06&  0.06$\pm$0.10& 0.00$\pm$0.07&--0.07$\pm$0.10&  0.01$\pm$0.04\\
HD 85301& 5640& 4.44& 0.05 &-0.05$\pm$0.12& --0.03$\pm$0.04& --0.04$\pm$0.10& 0.02$\pm$0.12& 0.01$\pm$0.08&  0.19$\pm$0.04\\
HD 87359*& 5645& 4.40&--0.07 & 0.01$\pm$0.10&  0.10$\pm$0.06&  0.04$\pm$0.10& 0.08$\pm$0.07& 0.01$\pm$0.12&  0.09$\pm$0.06\\
HD 87836& 5684& 4.05& 0.16 & 0.22$\pm$0.12&  0.09$\pm$0.05&  0.13$\pm$0.10& 0.02$\pm$0.10& 0.01$\pm$0.10&  0.03$\pm$0.05\\
HD 90681& 5950& 4.26& 0.16 &--0.02$\pm$0.12&  0.04$\pm$0.06&  0.05$\pm$0.07&--0.04$\pm$0.12&--0.04$\pm$0.10& --0.02$\pm$0.05\\
HD 90722& 5677& 4.12& 0.22 & 0.08$\pm$0.12&  0.04$\pm$0.05&  0.09$\pm$0.10&--0.04$\pm$0.10&--0.02$\pm$0.10& --0.05$\pm$0.03\\
HD 92788*& 5694& 4.26& 0.13 & 0.15$\pm$0.15&  0.07$\pm$0.10&  0.15$\pm$0.10& 0.06$\pm$0.07&--0.04$\pm$0.15&  0.04$\pm$0.05\\
HD 97645& 6127& 4.10& 0.14 &--0.02$\pm$0.15&  0.02$\pm$0.04&  0.09$\pm$0.10&--0.02$\pm$0.12&--0.02$\pm$0.12&  0.06$\pm$0.05\\
HD 98618*& 5727& 4.27&--0.11 & 0.07$\pm$0.08&  0.08$\pm$0.06&  0.08$\pm$0.07& 0.11$\pm$0.08& 0.00$\pm$0.12&  0.13$\pm$0.04\\
HD 98736*& 5276& 4.40& 0.29 & 0.17$\pm$0.15& --0.03$\pm$0.07&  0.06$\pm$0.10&--0.04$\pm$0.10&--0.08$\pm$0.15& --0.11$\pm$0.08\\
HD 99491& 5537& 4.40& 0.25 & 0.15$\pm$0.12&  0.04$\pm$0.07&  0.14$\pm$0.10&--0.07$\pm$0.12& 0.03$\pm$0.12& --0.06$\pm$0.06\\
HD 99492*& 5006& 4.56& 0.20 & 0.22$\pm$0.12&  0.00$\pm$0.09&  0.17$\pm$0.15&--0.05$\pm$0.12&--0.03$\pm$0.15& --0.03$\pm$0.06\\
HD100069& 5796& 3.74&--0.03 & 0.08$\pm$0.15&  0.11$\pm$0.08&  0.13$\pm$0.10& 0.06$\pm$0.08& 0.06$\pm$0.12&  0.09$\pm$0.04\\
HD105631& 5391& 4.47& 0.05 &--0.01$\pm$0.12&  0.04$\pm$0.06&  0.03$\pm$0.08& 0.03$\pm$0.08& 0.02$\pm$0.10&  0.10$\pm$0.04\\
HD106116& 5665& 4.30& 0.03 & 0.10$\pm$0.10&  0.05$\pm$0.06&  0.11$\pm$0.08& 0.00$\pm$0.08& 0.04$\pm$0.10&  0.05$\pm$0.03\\
HD106156& 5449& 4.45& 0.10 & 0.04$\pm$0.12&  0.04$\pm$0.08&  0.05$\pm$0.15& 0.02$\pm$0.12& 0.01$\pm$0.08&  0.01$\pm$0.04\\
HD108942& 5882& 4.27& 0.20 & 0.08$\pm$0.15&  0.00$\pm$0.08&  0.04$\pm$0.10&--0.09$\pm$0.10& 0.02$\pm$0.10& --0.05$\pm$0.06\\
HD114174& 5728& 4.30&--0.06 & 0.09$\pm$0.10&  0.08$\pm$0.07&  0.11$\pm$0.10& 0.02$\pm$0.10& 0.07$\pm$0.10&  0.28$\pm$0.06\\
HD116321& 6292& 3.66& 0.10 & 0.18$\pm$0.20&  0.13$\pm$0.04&  0.20$\pm$0.18& 0.06$\pm$0.12&--0.01$\pm$0.12&  0.17$\pm$0.08\\
HD136618& 5805& 3.56& 0.14 & 0.18$\pm$0.18&  0.09$\pm$0.05&  0.10$\pm$0.12&--0.03$\pm$0.10&--0.02$\pm$0.10&  0.02$\pm$0.05\\
HD145675*& 5312& 4.37& 0.33 & 0.18$\pm$0.15&  0.06$\pm$0.04&  0.07$\pm$0.10&--0.02$\pm$0.09&--0.03$\pm$0.10& --0.11$\pm$0.06\\
HD147231& 5594& 4.30&--0.14 & 0.09$\pm$0.15&  0.22$\pm$0.09&  0.09$\pm$0.15& 0.07$\pm$0.10& 0.20$\pm$0.12&  0.05$\pm$0.06\\
HD159222& 5815& 4.30& 0.00 & 0.05$\pm$0.10&  0.05$\pm$0.03&  0.05$\pm$0.10& 0.04$\pm$0.10&--0.04$\pm$0.12&  0.08$\pm$0.06\\
HD190360*& 5424& 4.21&--0.01 & 0.11$\pm$0.12&  0.17$\pm$0.08&  0.18$\pm$0.10& 0.20$\pm$0.09& 0.08$\pm$0.12&  0.05$\pm$0.05\\
\hline                                                                                            
\end{tabular}                                                                               
\end{table*}

To estimate the difference in the results between the NLTE and LTE 
approximations, for four representative stars of our sample, we determined the 
LTE abundances of the studied elements using the same set of lines and the same 
atomic parameters. The results are given in Table\,\ref{dNLTE}, where the LTE [X/Fe] is 
reported as well as the difference NLTE--LTE abundance ($\Delta$NLTE in the Table).
The choice of the four stars was driven by the desire to span the stellar parameter space.
In Fig.\,\ref{fig:hr} the selected stars are highlighted (red open circle) over the complete sample in the 
\teff, \logg\ space.

\begin{table*}
\caption{LTE abundances for selected stars.}
\label{dNLTE}
\begin{tabular}{rccrccl|ccl|ccccccc}
\hline
Star&T$_{\rm eff}$, K&log g&[Fe/H]&[Na/Fe]&$\sigma_{\rm LTE}$&$\Delta$NLTE &[Mg/Fe]&$\sigma_{\rm LTE}$&$\Delta$NLTE &[Al/Fe]&$\sigma_{\rm LTE}$&$\Delta$NLTE\\
\hline                                                                                        
\hline                                                                                        
 HD98618& 5727& 4.27&        --0.11&   0.09$\pm$0.09&    0.04&      --0.02 &   0.12$\pm$0.07&    0.04&       --0.04 & 0.13$\pm$0.09  & 0.06   & --0.05      \\
HD106156& 5449& 4.45&         0.10&   0.06$\pm$0.14&    0.07&       --0.02 &   0.08$\pm$0.10&    0.06&       --0.04 & 0.12$\pm$0.16  & 0.06   & --0.07      \\
HD116321& 6292& 3.66&         0.10&   0.22$\pm$0.25&    0.15&       --0.04 &   0.15$\pm$0.06&    0.04&       --0.02 & 0.26$\pm$0.19  & 0.04   & --0.06      \\
HD145675& 5312& 4.37&         0.33&   0.21$\pm$0.16&    0.04&       --0.03 &   0.10$\pm$0.06&    0.05&       --0.04 & 0.12$\pm$0.11  & 0.03   & --0.05      \\
\hline                                                                                            
Star&T$_{\rm eff}$, K&log g&[Fe/H]&[S/Fe]&$\sigma_{\rm LTE}$&$\Delta$NLTE & [K/Fe]&$\sigma_{\rm LTE}$&$\Delta$NLTE &[Sr/Fe]&$\sigma_{\rm LTE}$&$\Delta$NLTE\\
\hline                                                                                                                        
\hline                                                                                                                        
 HD98618& 5727& 4.27&        --0.11&  0.14$\pm$0.11&    0.07&      --0.03 &  0.18$\pm$0.15 & 0.09   &      --0.18 &   0.41$\pm$0.05&    0.03&       --0.28\\
HD106156& 5449& 4.45&         0.10&  0.04$\pm$0.14&    0.04&       --0.02 &  0.18$\pm$0.14 & 0.12   &       --0.17 &   0.25$\pm$0.05&    0.03&       --0.24\\
HD116321& 6292& 3.66&         0.10&  0.14$\pm$0.18&    0.13&       --0.08 &  0.20$\pm$0.18 & 0.13   &       --0.21 &   0.62$\pm$0.09&    0.03&       --0.45\\
HD145675& 5312& 4.37&         0.33&  0.00$\pm$0.10&    0.04&       --0.02 &  0.11$\pm$0.14 & 0.10   &       --0.14 &   0.07$\pm$0.07&    0.04&       --0.18\\
\hline                                                                                                                    
\end{tabular}                                                                               
Notes: $\sigma_{\rm LTE}$ -- the additional error introduced by the use of LTE approximation.
\end{table*}

\subsection{{Na}{\sc I}} 
The NLTE sodium model atom was first described in \citet{KorMish99} and later updated by \citet[see][]{Dobrovolskas14}. This model provides the NLTE solar sodium abundance A(Na)=6.25, derived from optical lines and reported in Table\,\ref{sun}. 
The oscillator strengths for the 1074.6--1267.9\,nm lines were taken from  
\citet{Wiese69}, for the 2205.6 and 2208.3\,nm lines from  \citet{McEachran83},
and for the 2334.8--2337.9\,nm lines from the NIST database \citep{Reader12}.

The theoretical NLTE synthesis of the 996.1\,nm line cannot reproduce the observed solar spectrum, as 
well as the 2145.2\,nm line. The problem is that these observed lines are weaker than in the theoretical synthesis. The estimated NLTE effects cannot explain such a difference, therefore those lines were excluded from the analysis. 

For all the lines (except one at 1267.9\,nm) we took van der Waals parameter: $\Gamma_{vw}$ = --7.06 
(hereafter we use $\Gamma_{vw}$ as a $\log$ of the line FWHM per perturber 
at $T=10000$\, K in $cgs$ units). This value is slightly larger than the estimated value from Uns{\"o}ld's formula  \citep{Unsold55} but it provides a better fit with observed solar profile. 

For the atomic model of sodium, as well as for the other models considered 
here, we must make the following remark. Correction of $\Gamma_ {vw}$, 
if necessary, is performed only for lines with developed wings. For such lines, 
the influence from the accepted value of $\Gamma_ {vw}$ is higher than from the 
uncertainty of the oscillator strength (the latter possibility, of course, is 
not excluded). Such line profiles cannot be adjusted only by changing the 
element abundance or the value of \loggf.

Below we list some notes on individual lines when comparing the NLTE theoretical synthesis to the observed
solar spectrum.
\begin{itemize} 
\item The synthesis of the line at 1074.644\,nm reproduces very well the observed solar spectrum.
\item The lines at 1083.484, 1083.484, 1083.490\,nm give abundances by about 0.1\,dex lower 
than our adopted solar sodium abundance. 
\item The lines at 1267.917, 1267.917, and 1267.922\,nm give abundance by 0.05\,dex
higher than our NLTE solar sodium abundance. $\Gamma_{vw}$ for these lines was  taken from \citet{Barklem00}. 
\item For three lines (2205.640, 2208.366 and 2334.837\,nm) the agreement between the NLTE theoretical synthesis and the solar observed spectrum is very good, while in the two lines at 2337.896 and 2337.914\,nm a higher solar abundance (by 0.04\,dex) would be needed to reproduce the solar observed spectrum.
\item All the lines are formed practically in LTE (EWs vary in the range from --2\% to +4\%). 
\item The exception is multiplet 2334.8--2337.9\,nm whose lines in NLTE are strengthened  by about 7--15\%. 
\end{itemize}

It is possible that a deviation within 0.1\,dex of the sodium content obtained from 
the individual lines listed above from the adopted solar sodium abundance of Table\,\ref{sun} 
may be due to inaccuracy of our adopted  oscillator strengths. 
An investigation of this issue  
is beyond the scope of this work. Similar conclusion can also be 
applied to some lines of other ions, discussed below.

Our NLTE abundance of sodium in the program stars differ from LTE abundances obtained 
by \citet{Caffau19a} from 0 to 0.25\,dex. On average, our NLTE abundance is 
0.09\,dex higher than the LTE abundance of \citet{Caffau19a}. We should stress that this comparison is not 
just as NLTE -- LTE difference because the lines used are not the same and the atomic parameters used neither.

The relative abundance [Na/Fe] of our sample stars shows a weak 
trend with metallicity, the effect is an increase with increasing metallicity (the range is from 
--0.08 to 0.22 dex).  A non parametric test with Kendall's $\tau$ provides in fact a correlation probability of 99\%. The same test on the subset of 1\,679 stars with Na measurements shows also  a
99\% probability of correlation. If we divide the two samples in 0.1\,dex metallicity bins
the GIANO sample has $\sigma$ in each bin that ranges from 0.007\,dex to 0.080\,dex. For the APOGEE sample it ranges from 0.27\,dex to 0.36\,dex. The mean abundance for the 40 stars is [Na/Fe] = +0.09 $\pm$ 0.07. 
It should be noted that the abundance from IR lines agrees well with abundance derived 
from optical lines (the mean deviation is only +0.03\,dex). Individual 
deviations do not exceed $\pm 0.06$\,dex.

\begin{figure}
\includegraphics[width=0.98\columnwidth,clip=true]{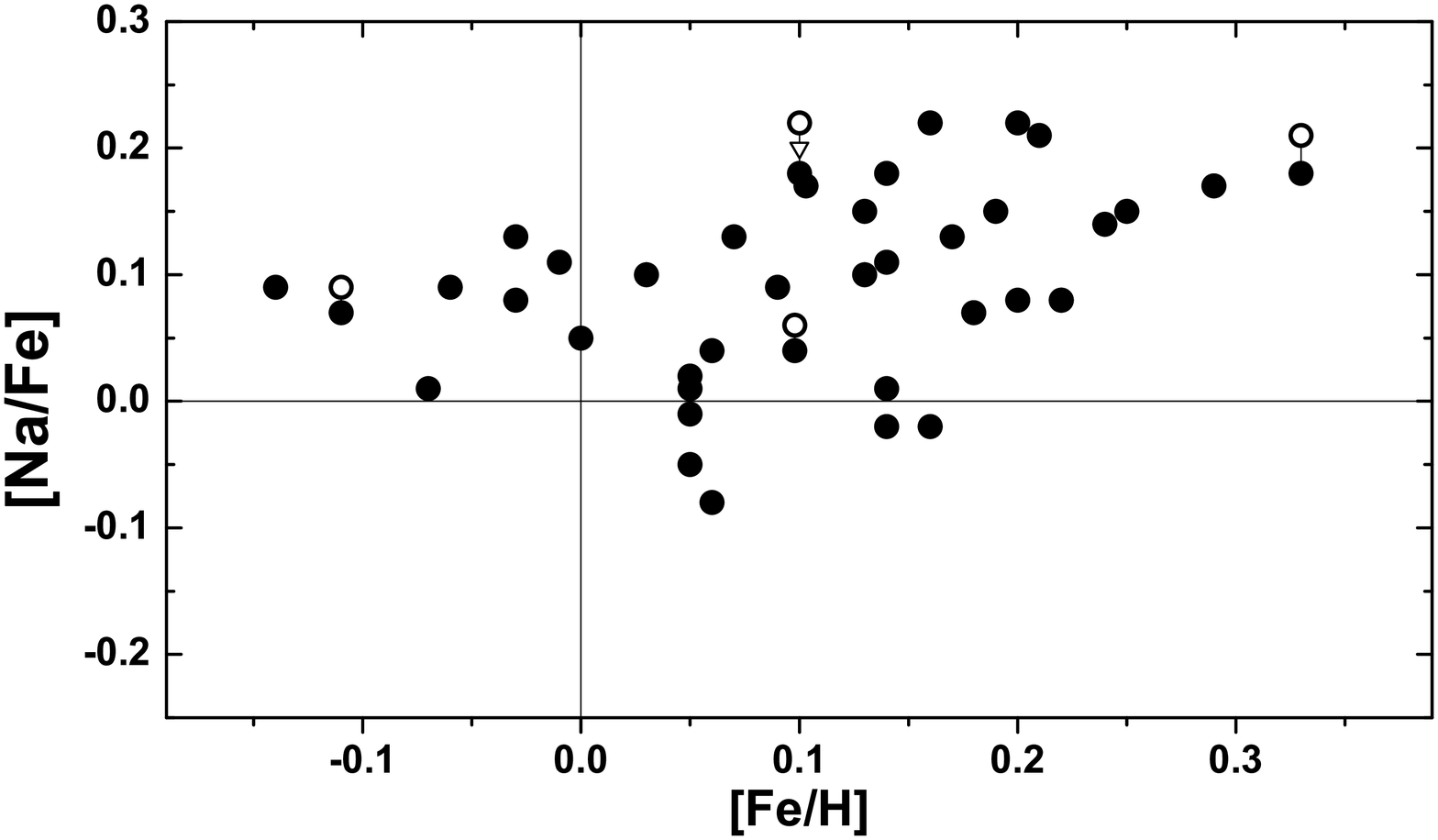}
\includegraphics[width=0.98\columnwidth,clip=true]{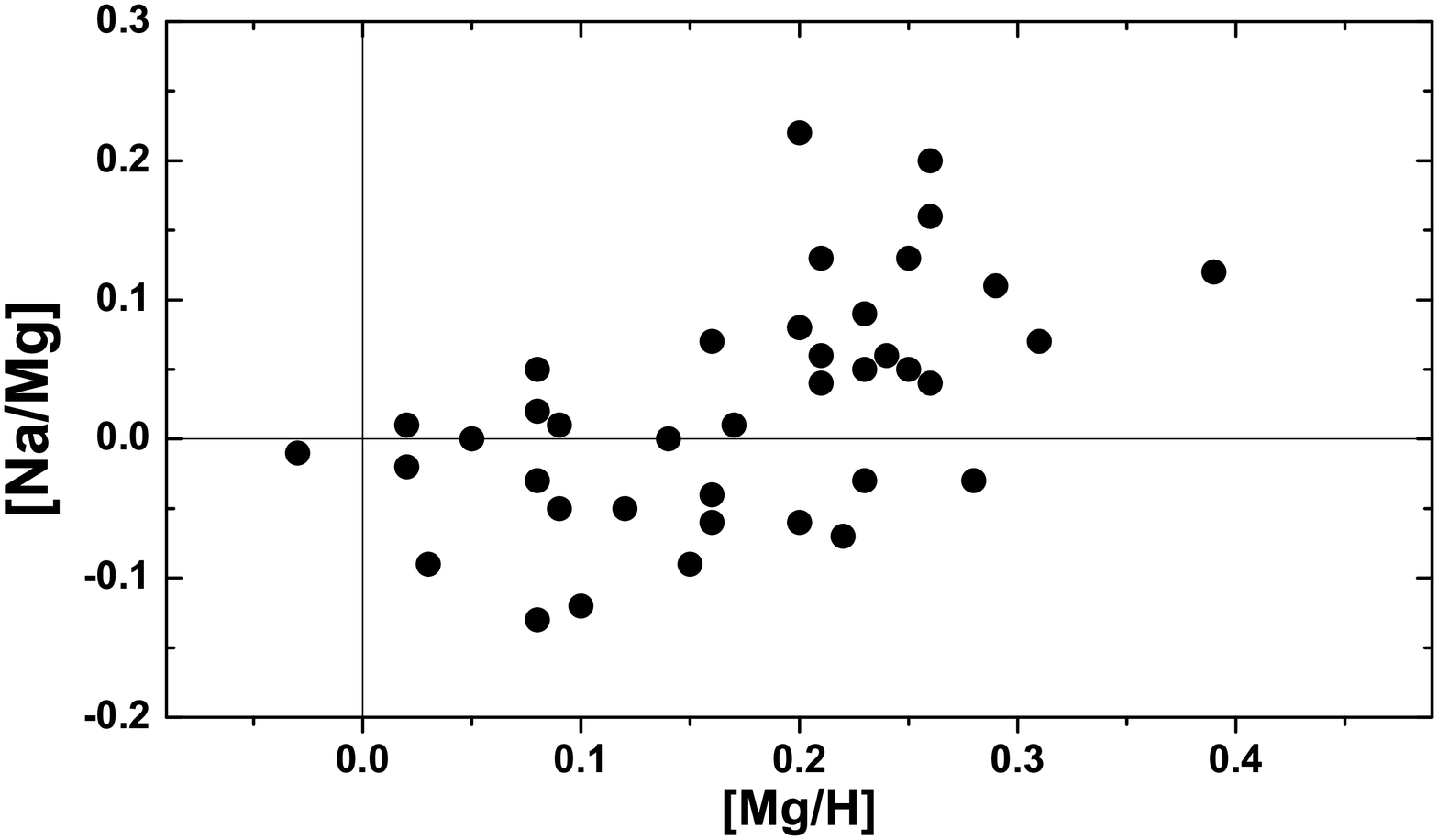}
\caption{Relative sodium abundances ([Na/Fe] and [Na/Mg]) as a function 
of [Fe/H] ($upper~panel$) and [Mg/H] ($lower~panel$). The NLTE abundances 
are indicated by $filled~circles$. For comparison we show also the LTE sodium 
abundances for the four selected stars ($open~circles$). 
The arrow lines connect the LTE and NLTE abundance values for the same stars. 
The numerical values of LTE abundances are listed in Table\,\ref{dNLTE}.}
\label{na_fe}
\end{figure}

In Appendix (Fig. \ref{linefit1}) we show some examples of the observed and synthetic 
NLTE and LTE profiles of two sodium lines in the solar and stellar observed spectra. 
Fig.\,\ref{na_fe} shows the derived NLTE sodium abundances for our sample of stars 
vs. [Fe/H] and [Mg/H] ($\alpha$-element). 
The A(Fe) determination is from \citet{Caffau19a}, so it is in LTE
while the A(Mg) is the value here derived in NLTE.
We understand the inconsistency in mixing LTE and NLTE abundances, but still we think useful
to provide, for each element, [X/Fe] versus [Fe/H], which is what usually presented in the literature.

The NLTE effects for Na in this stellar parameter space are well within the uncertainties, being for the four
representative stars at maximum --0.04\,dex. 

\subsection{{Mg}{\sc I}}
The NLTE magnesium (\ion{Mg}{I}) atomic model was described in 
\citet{Mishenina04} and later updated by \citet{Cerniauskas17}.
Our NLTE solar magnesium abundance from optical lines is A(Mg)=7.54 (see Table\,\ref{sun}). 
The oscillator strength for the 1081.12\,nm line was taken from
\citet{Kurucz75}; for the 1203.9--1208.3\,nm lines -- from \citet{Chang90}, 
and for other lines -- from NIST database (the latter uses mainly the results of 
calculations by \citealt{Butler93}).
The van der Waals parameters were calculated using Uns{\"o}ld's 
formula and then corrected with the help of lines from solar spectrum. For line at 1182.8\,nm 
we used the value reported in \citet{Barklem00}. NLTE effects lead to the 
strengthening of the line in vicinity of the core, while the
wings remain practically unchanged.  
The lines at 982.1, 982.8, 998.3, 998.6\,nm are formed practically in LTE. To fit the synthetic line
profiles to the solar observed ones, by keeping our adopted solar abundance, 
one needs to 
increase the oscillator strengths of these 
lines by about 0.08\,dex, which is of the same order of magnitude of the line-to-line scatter
for the Mg investigation (see Table\,\ref{ab}). 

Some comments on the lines here below about the solar study of the Mg lines.
\begin{itemize}
\item The feature at 1081.1\,nm is formed by 8 components. The NLTE corrections are not negligible 
(exceed 5\% in EW). The wings of this feature can be adjusted if $\Gamma_{vw}$ is 
decreased from --6.68 \citep{Barklem00} to --6.82 (the value derived from the solar spectrum).  
\item The lines at 1095.3, 1095.7 and 1096.5\,nm are formed practically in LTE.  
\item For line at 1182.8\,nm the NLTE correction in EW achieves 6\%. 
\item The line at 1203.9\,nm is formed practically in LTE.  
\item The NLTE effect for the 1208.3\,nm feature is modest (about 5\% in EW). This feature is a 
blend of two lines. The wing of this blend can be adjusted if $\Gamma_{vw}$ 
is decreased from --6.98 \citep{Barklem00} to --7.12 (from the solar spectrum).  
\item The lines at 1241.7, 1242.3, 1243.3\,nm are formed practically in LTE.  
\item The lines at 1502.4, 1504.0, 1504.7\,nm show very small NLTE effects.
\item The lines at 1574.0, 1574.8, 1576.5\,nm also show very small NLTE effects. 
\item For the 1710.8\,nm line, the NLTE effects are about 5\% in EW.  
\end{itemize}

\begin{figure}
\includegraphics[width=0.98\columnwidth,clip=true]{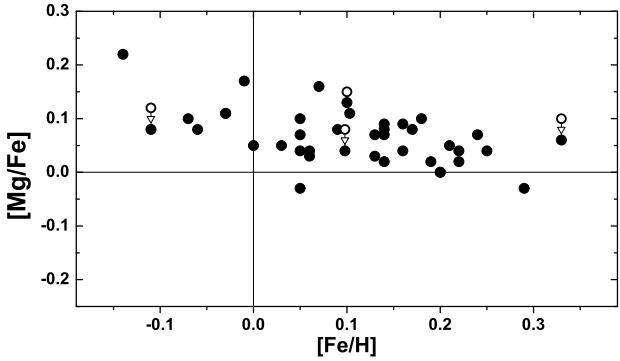}
\caption{Same as Fig.~\ref{na_fe} (upper panel) but for magnesium.}
\label{mg_fe}
\end{figure}

The NLTE correction for Mg for the four selected stars is small, up to --0.04\,dex.
In Appendix (Fig. \ref{linefit1}) we show NLTE and LTE profiles of two magnesium lines in the solar and stellar 
spectra. Fig. \ref{mg_fe} is the same as Fig. \ref{na_fe} but for magnesium. 

From our data one can infer a decrease of the [Mg/Fe] ratio with increasing
metallicity with a probability of over 99\%.
Such a decrease is also clear from the APOGEE data,
however the decrease is not as steep as in our data.
Our [Mg/Fe] ratios are higher than the APOGEE data 
by about 0.08\,dex for the
lowest metallicity bin and converge towards the APOGEE values at higher metallicities.

\subsection{{Al}\,{\sc I}} 
The aluminum NLTE atomic model is described in detail in \citet{Andrievsky08} 
and \citet{Caffau19b}. Our adopted NLTE solar abundance is A(Al)=6.43 (see 
Table\,\ref{sun}). 
The van der Waals  parameters for the IR lines were obtained from 
Uns{\"o}ld's formula. For some lines the van der Waals parameters were 
corrected using the solar spectrum. Some details on individual lines while 
comparing theoretical NLTE profile to the observed solar spectrum are given 
below.
\begin{itemize}
\item The line at 1087.297\,nm (NIST \loggf) is very 
well modeled by the synthetic spectrum. This line is formed in LTE. 
\item The line at 1089.174\,nm (NIST \loggf) 
cannot be reproduced by theoretical synthesis.  Probably this line is a blend, 
because its \loggf listed in NIST has an accuracy better than 10\%.
The estimated NLTE correction is negligible for this line. 
\item The theoretical NLTE profile of line at 1312.341\,nm 
(\loggf are from \citealt{Wiese69})
reproduces very well the observed solar profile, once the $\Gamma_{vw}$ is adjusted. 
The LTE profile is wider in the wings but shallower in the core compared to observed profile. 
The NLTE EW in the solar spectrum appears to be somewhat larger but not 
very significantly (3--5\%) with respect to the LTE EW. For the stars this difference can be larger (up to 
13 \%, which corresponds to 0.15 dex in abundance).
\item The line at 1315.075\,nm (\loggf are from \citealt{Wiese69} and $\Gamma_{vw}$ 
was derived to fit the solar spectrum) is blended in the solar spectrum 
with a telluric line. However, the profile of this line in the spectra of the studied stars
is well described with the same abundance as derived from the 1312.3\,nm line.
The value of the $\Gamma_{vw}$ is taken the same as for the line 1312.3\,nm of the same multiplet.
Therefore, we can assume that parameters of this line are fairly reliable.
\item Synthetic profile of the line at 1671.896\,nm (NIST \loggf and $\Gamma_{vw}$ was
optimised to reproduce the solar spectrum) is weaker than the observed one. 
This result cannot be attributed to an
inaccuracy in \loggf value, since  
the accuracy on the oscillator strength is better than 3\%. 
The core of the line profile in LTE is shallower than what is
calculated in NLTE. In any case NLTE effects are small for this line. 
\item The central part of the line at 1675.056\,nm (NIST \loggf and $\Gamma_{vw}$ 
was adjusted to fit the solar spectrum) is blended with a telluric line in the solar spectrum. 
This does not allow us to estimate the accuracy of the value $\log\,gf$.
In the stellar spectra the synthetic profile of this line is systematically
weaker than observed one. The difference is not the result of the \loggf inaccuracy, since it is better than 3\%.
\item The line at 1676.336\,nm  (NIST \loggf and $\Gamma_{vw}$ was adjusted to fit the solar spectrum) is described very well in the solar spectrum. For the stellar spectra the difference LTE and NLTE in EW can 
achieve up to 4\% (which corresponds to 0.05 dex in abundance).
\item The line at 2109.303\,nm  ($\log\,gf$ is from \citealt{Kurucz75}) is 
This does not allow us to check the value of $\Gamma_{vw}$. 
Therefore, we use the parameter $\Gamma_{vw}$ the same as for the line 2116.376\,nm
of the same multiplet.
For the stars, the abundance derived from this line, agrees with abundance derived from the line at 2116.3\,nm. LTE profile of this line is shallower than in the observed spectrum. For the Sun, the NLTE EW is slightly larger but insignificantly (3--5 \%) than the LTE value. For the stellar spectra this difference can achieve up to 12 \%. 
\item The NLTE synthetic profile of the line at 2116.376\,nm 
(\loggf is from \citealt{Kurucz75} and $\Gamma_{vw}$ was derived from the solar spectrum) 
reproduces very well the solar observed spectrum.
\end{itemize}

In order to derive aluminum abundance in the dwarf stars, using GIANO spectra, we used the 
following lines: 1087.2, 1312.3, 1315.0, 1671.8, 1676.3, 2109.3, 2116.3\,nm. 
The abundance of aluminum derived from the IR lines is practically the same as from the optical 
lines at 669.6, 669.8, 736.1, 736.2, 783.5, 783.6, 877.2, 877.3\,nm. The difference is no more 
than 0.04\,dex. Our NLTE aluminum abundance differs from LTE abundance derived by 
\citet{Caffau19a} up to $\pm$0.15 dex. The mean difference is about 0.05\,dex 
(our value is lower) but this difference is probably  due to 
the differences in adopted atomic data and lines used.
 
[Al/Fe] does not show any dependence on metallicity. 
This value is distributed in the range from --0.05 to 0.20\,dex with a mean value 
from 40 stars [Al/Fe] = +0.09 $\pm$ 0.05. 

\begin{figure}
\includegraphics[width=0.98\columnwidth,clip=true]{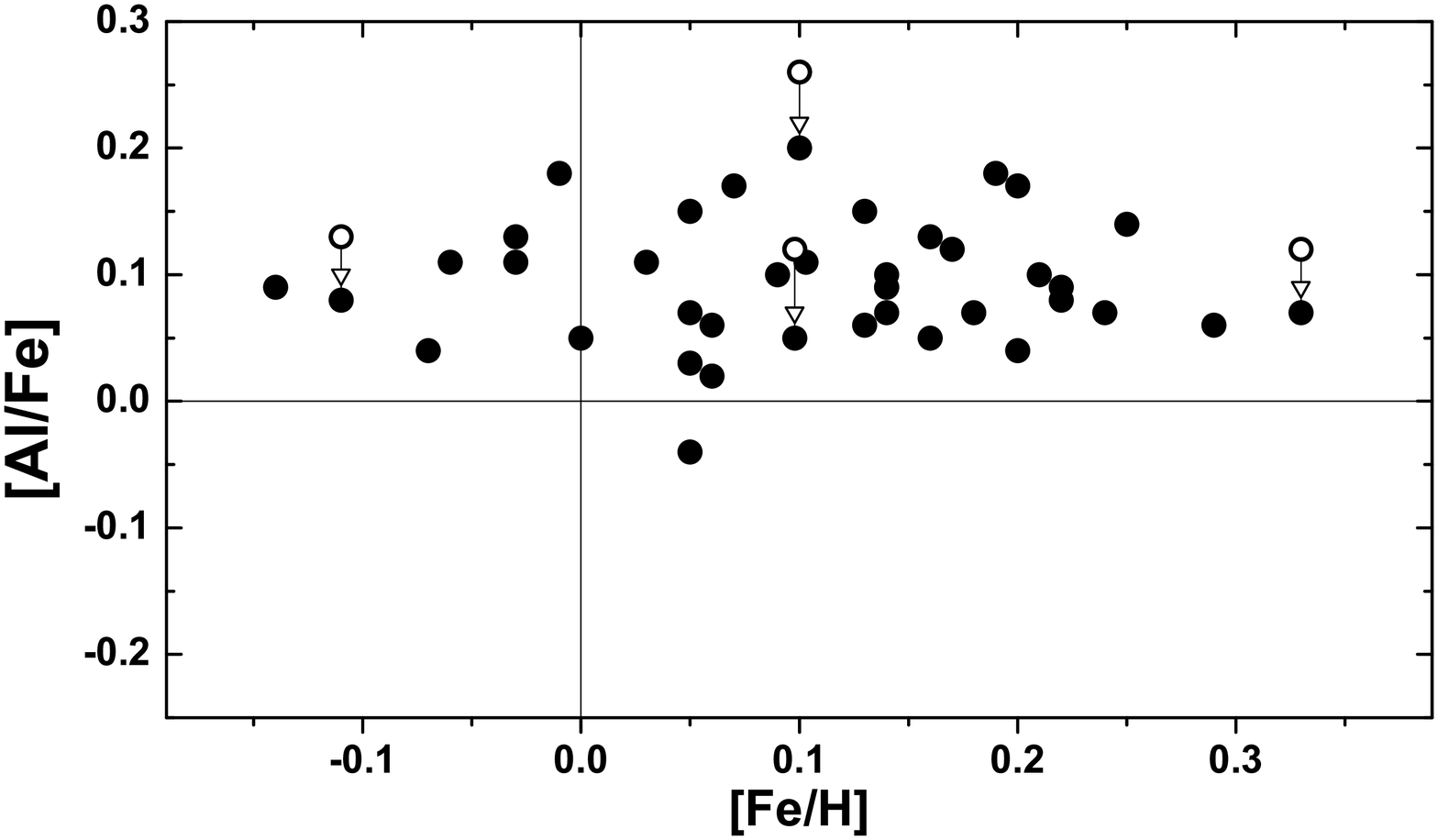}
\includegraphics[width=0.98\columnwidth,clip=true]{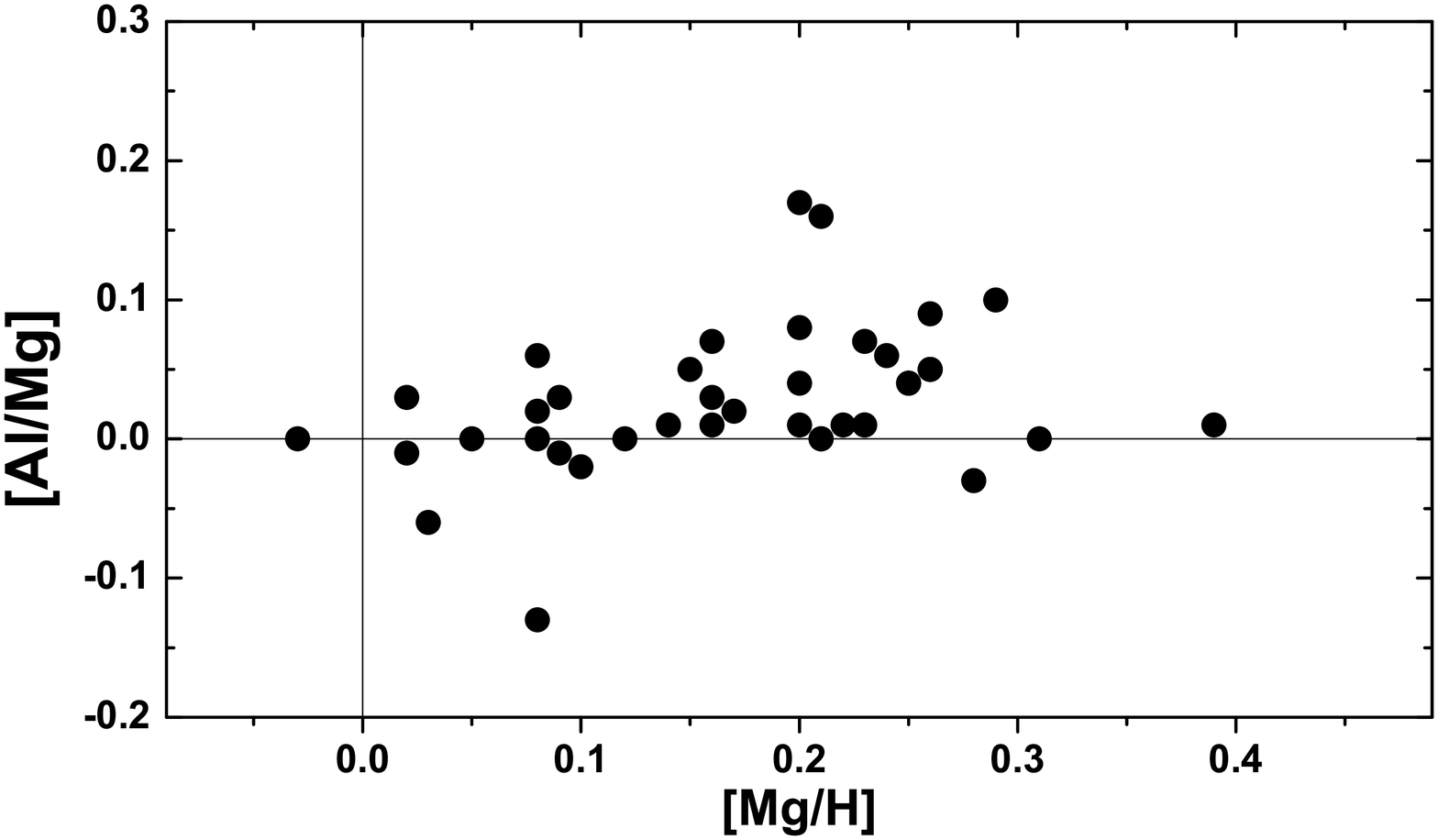}
\caption{Same as Fig.~\ref{na_fe} but for aluminum.}
\label{al_fe}
\end{figure}

The NLTE correction we derived for the four selected stars is small, on average about --0.06\,dex.
In Appendix (Fig. \ref{linefit1}) we show  NLTE and LTE profiles of two aluminum lines in the solar and stellar 
spectra. Fig.\,\ref{al_fe} is the same as Fig. \ref{na_fe} but for aluminum.

\subsection{{S}\,{\sc I}}

The sulfur NLTE model is described in detail in \citet{Korotin09}. For the solar 
sulfur abundance this model, used in the optical wavelength range, provides 
A(S)=7.16 (see Table\,\ref{sun}). 
For the lines at 921.2--923.7\,nm, the oscillator strengths we applied are from 
\citet{Biemont93}, as well as for the lines at 2250.0--2270.0\,nm. 
The synthetic profiles of the lines at 1045.5, 1045.6, 1045.9\,nm are in good 
agreement with the solar spectrum when using values \loggf  from \citet{Wiese69}.
Van der Waals parameters for these lines are from \citet{Barklem00}.

For the lines in the region 1540.0--1542.3\,nm, the oscillator strengths were 
taken from two sources: \citet{Biemont93} and NIST. 
It should be noted that the \loggf from NIST are lower by 0.18\,dex with 
respect to the data of \citet {Biemont93}.  The same is valid for the lines of 
the multiplet at 1546.9--1547.9\,nm, where  the difference between the \loggf 
is 0.16\,dex. 
The lines at 1540.0, 1540.3 \,nm are blended with telluric lines 
in the solar spectrum, therefore we cannot use them to check their parameters. 
The lines at 1542.3, 1546.9, 1547.5, 1547.8\,nm are clean, and we found that 
their \loggf values can be decreased by 0.1\,dex (with respect to the data 
of \citealt{Biemont93}) to achieve a good agreement  with the observed  profiles 
in the solar spectrum. 
It is important to note that these lines are weak in the solar spectrum
(residual intensity is larger than 0.85), therefore, the influence of the accepted values 
of \loggf on their calculated strength is predominant.
Thus, we one can conclude that, most likely, the real values of \loggf for mentioned lines 
appear to be between the data of \citet {Biemont93} and NIST. The NIST accuracy for 
these oscillator strengths are $\geq 25$\%, and they are listed in Table\,\ref{lin}.
The lines at 2250.7, 2251.9\,nm are reproduced with a reasonable agreement in the solar 
spectrum but the continuum in this region is not quite reliable. The lines at 2255.2, 
2256.3, 2257.5, 2264.4\,nm are slightly blended with telluric lines in the solar 
spectrum, and the line at 2270.7\,nm is very well reproduced by a synthetic profile. 

Actually, in our NLTE analysis of dwarf spectra we used the following lines: 
1045.5, 1045.6, 1045.9\,nm\ (principal lines for which NLTE corrections are 
moderate: EWs are up to 25\%, i.e. --0.3\,dex in abundance). The NLTE corrections increase 
as effective temperature increases. For most stars the corrections are within 0.15\,dex.
The lines at 1540.0--1547.9\,nm have the NLTE corrections up to 5\% in EW. 
For these lines, we derive a correction in abundance of +0.10 dex, 
which can be due to an inaccuracy of the oscillator strengths
(see discussion in the beginning of this subsection). 
The lines at 2250.7, 2251.9, 2255.2, 2257.5, 2256.3, 2264.4, 2270.7\,nm are 
formed almost in LTE (difference in EWs is about 3\%).

For our analysis in the optical range we used the following lines: 921.2, 
922.8, 923.7\,nm (large NLTE corrections: from $-$0.13 to $-$0.18 dex), 
869.3 and 869.4\,nm (corrections are rather small: from $-$0.05 to $-$0.03 dex), 
674.3, 674.8, 675.7, 605.2\,nm (lines are formed almost in LTE). 
The agreement between the abundances derived from GIANO spectra and the UVES and 
ESPaDOnS spectra is excellent (within the range of 0.04 dex).  Our NLTE sulfur abundances 
for individual stars differ from LTE values of \citet{Caffau19a} from --0.17 to 0.15 dex
and our mean NLTE abundance for all sample of stars is about 0.02\,dex 
lower than the  mean LTE value derived by \citet{Caffau19a}, but these comparisons
are  related to the difference in selected atomic data and lines used,  
not to NLTE effects.
The ratio [S/Fe] demonstrates a prominent increase with 
metallicity decrease (probabilty over 99\%). 
Such an increase is stastically apparent also in
the APOGEE data, however in our data it is considerably steeper, 
with [S/Fe] ratios higher than in the
APOGEE data  at the lowest metallicities and lower
at the highest metallicity, with a cross-over at about 0.1\,dex.
The ratios for individual stars are within the interval 
from --0.09 to 0.20\,dex. The mean value from 40 stars is [S/Fe]= +0.01$\pm$0.05.

\begin{figure}
\includegraphics[width=0.98\columnwidth,clip=true]{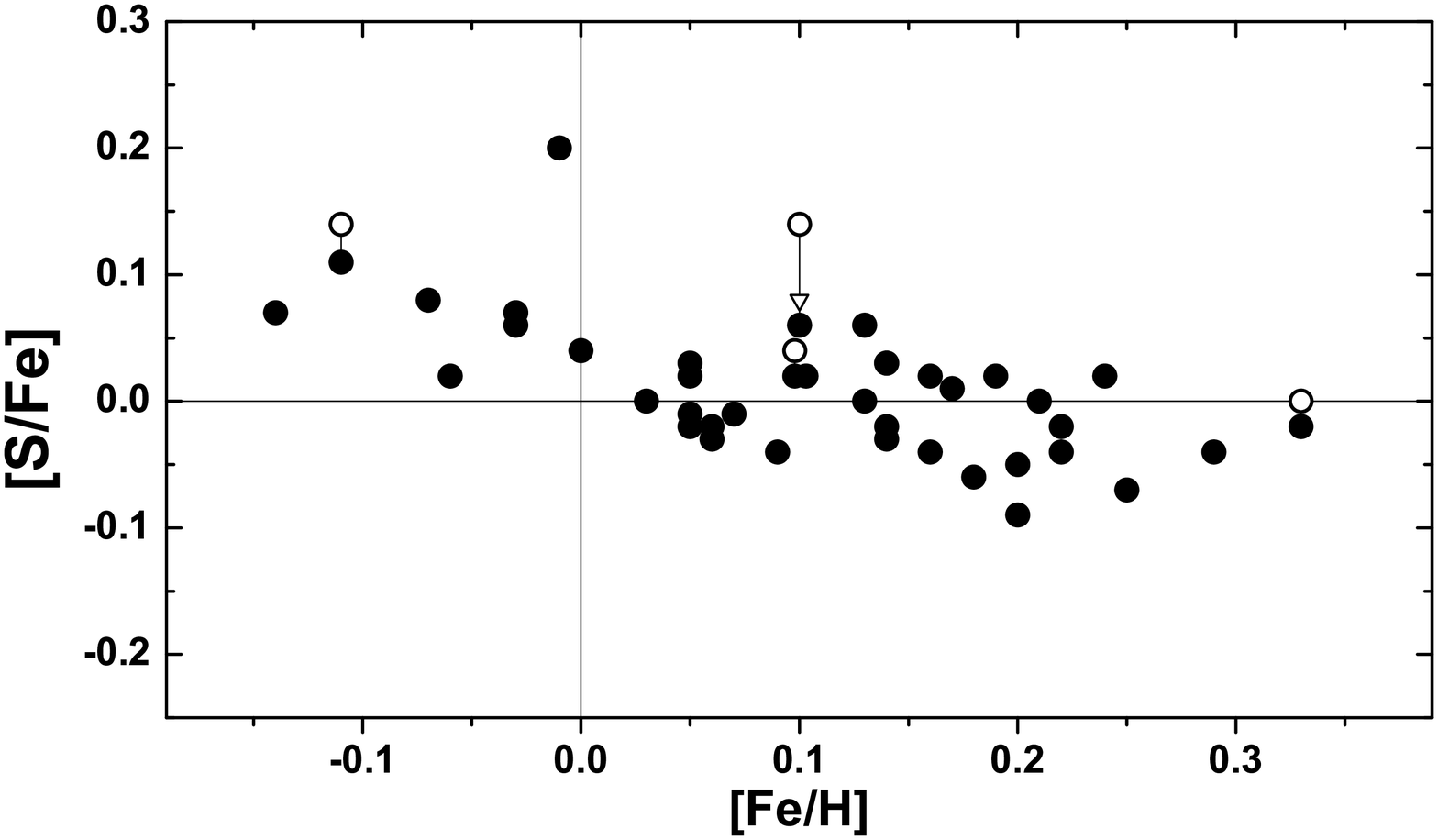}
\includegraphics[width=0.98\columnwidth,clip=true]{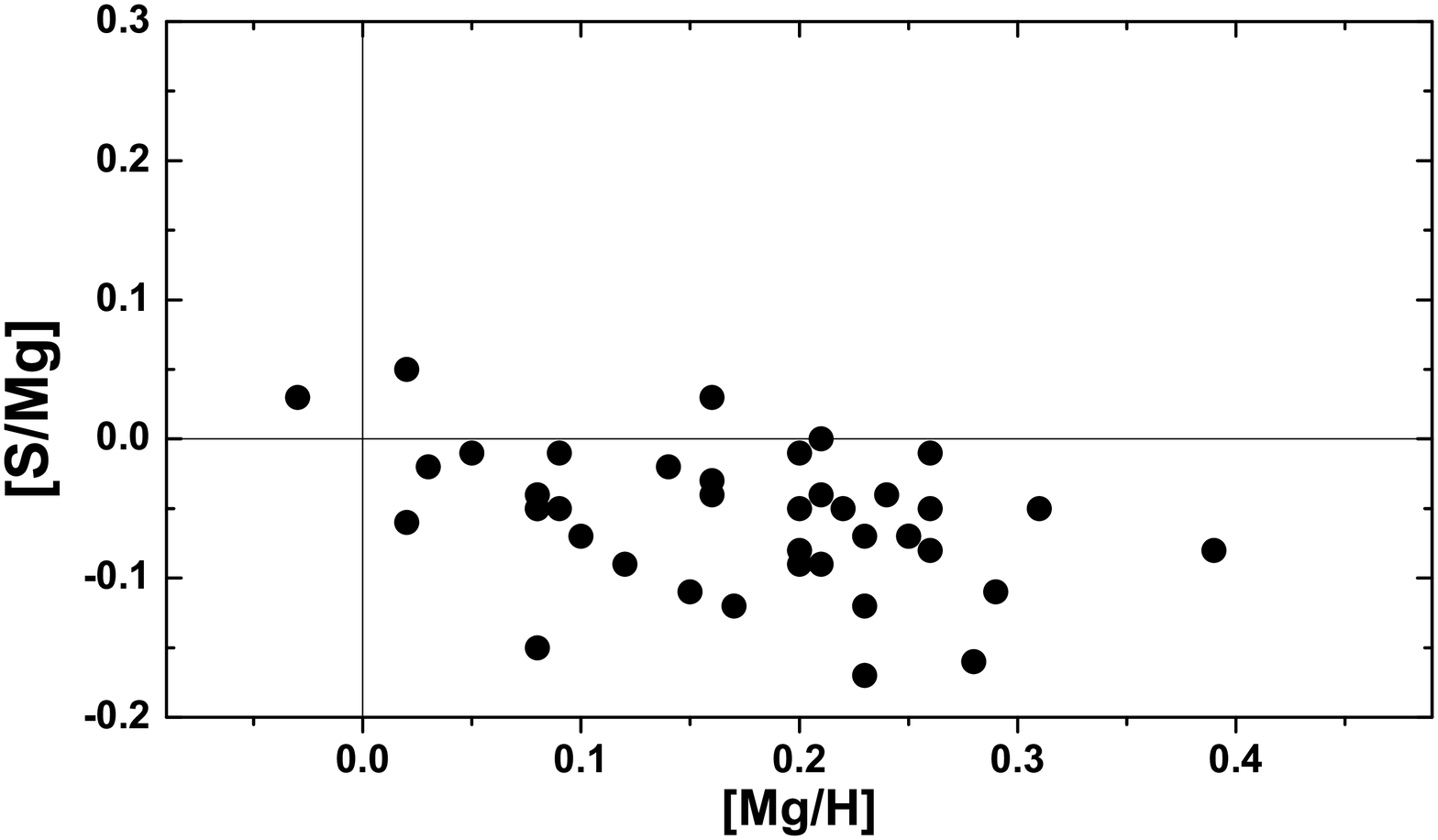}
\caption{Same as Fig.~\ref{na_fe} but for sulfur.}
\label{s_fe}
\end{figure}

The sulphur NLTE correction, we derived for the four selected stars, is negligible for the three 
cooler stars, for the hottest star it is --0.08\,dex, but anyway smaller than the line-to-line scatter.
In Appendix (Fig. \ref{linefit1}) we show  NLTE and LTE profiles of two sulfur lines in the solar and 
stellar spectra. Fig.\,\ref{s_fe} is the same as Fig. \ref{na_fe} but for sulfur.

\subsection{{K}\,{\sc I}} 
The potassium NLTE atomic model was described in \citet{Andrievsky10}.  
The basic atomic model was modified: photoionization cross 
sections for some levels were added following to the data from by
\citet{Zatsarinny10}; then, for the six lower levels collisional rates with 
hydrogen atoms were added \citep{Belyaev17}. 
The oscillator strengths for the lines at 1176.9 and 1177.2\,nm were taken 
from \citet{Safronova13}, for other lines -- from \citet{Wiese69}.
For the lines at 1176.9, 1177.2, 1243.2, 1252.2\,nm damping parameters 
$\Gamma_{vw}$ are from \citet{Barklem00}. For the lines at 1516.3 and 1516.8\,nm 
van der Waals parameters were calculated with the help of from Uns{\"o}ld's 
formula. Our solar potassium abundance derived from the optical range is A(K)=5.11 (see Table\,\ref{sun}). 
All potassium lines in the IR domain of the solar spectrum are 
strengthened due to NLTE effects (abundance corrections are from 0.03 to 0.18\,dex). 

There are some important details. In the stellar spectra the right wing 
of the 1176.9\,nm line is distorted by CN lines. In the solar spectrum these lines are 
separated, and potassium line is well reprouced by theoretical synthesis. A telluric absorption blends the 
1177.2\,nm line in the solar spectrum. The line at 1243.2\,nm is well reproduced by the synthetic profile.
The left wing of the 1252.2\,nm line in the stellar spectra is distorted by the presence of the 1252.181\,nm 
\ion{Cr}{I} line, but it is not a problem for the solar spectrum. The line at 1516.3\,nm is situated in 
a telluric absorption wing in the solar spectrum. In the stellar spectra the left wing 
of this line is distorted by the presence ot the 1516.265\,nm CN band. This line is formed in LTE. 
Similar situation is seen for the 1516.8\,nm line, which is also blended by a telluric 
line, and formed in LTE. The right wing of this line in the stellar spectra is 
distorted by the 1516.866\,nm CN band and the 1516.887\,nm \ion{Fe}{I} line. 

For the stellar spectra we used as a rule the 1176.9, 1177.2, 1243.2, 1252.2\,nm lines. 
The lines at 1516.3 and 1516.8\,nm were used as auxiliary lines (since they are 
partially blended with CN bands). In order to compare the abundances from 
GIANO to those derived from UVES and ESPaDOnS, 
we used the 766.4 and 769.8\,nm lines. We found that the 
differences in abundances are within 0.05\,dex. Only for one star, HD98736, the 
difference was 0.11\,dex (the abundance from IR lines is higher). The difference 
of the mean abundances from NLTE analysis and LTE analysis by \citet{Caffau19a} 
is up to 0.3\,dex; in this difference the NLTE effects play a part but the difference is also related to atomic data choice. 
[K/Fe] vs. [Fe/H] shows a small increase of the potassium 
abundance as metallicity decreases, this is statistically significant
(99.7\%). 
A similar increase is statistically apparent also in the APOGEE data, 
although impossible to detect visually, even after binning the
data in metallicity.The mean potassium abundance based 
on the 40 stars is [K/Fe] = +0.01$\pm$0.05.

\begin{figure}
\includegraphics[width=0.98\columnwidth,clip=true]{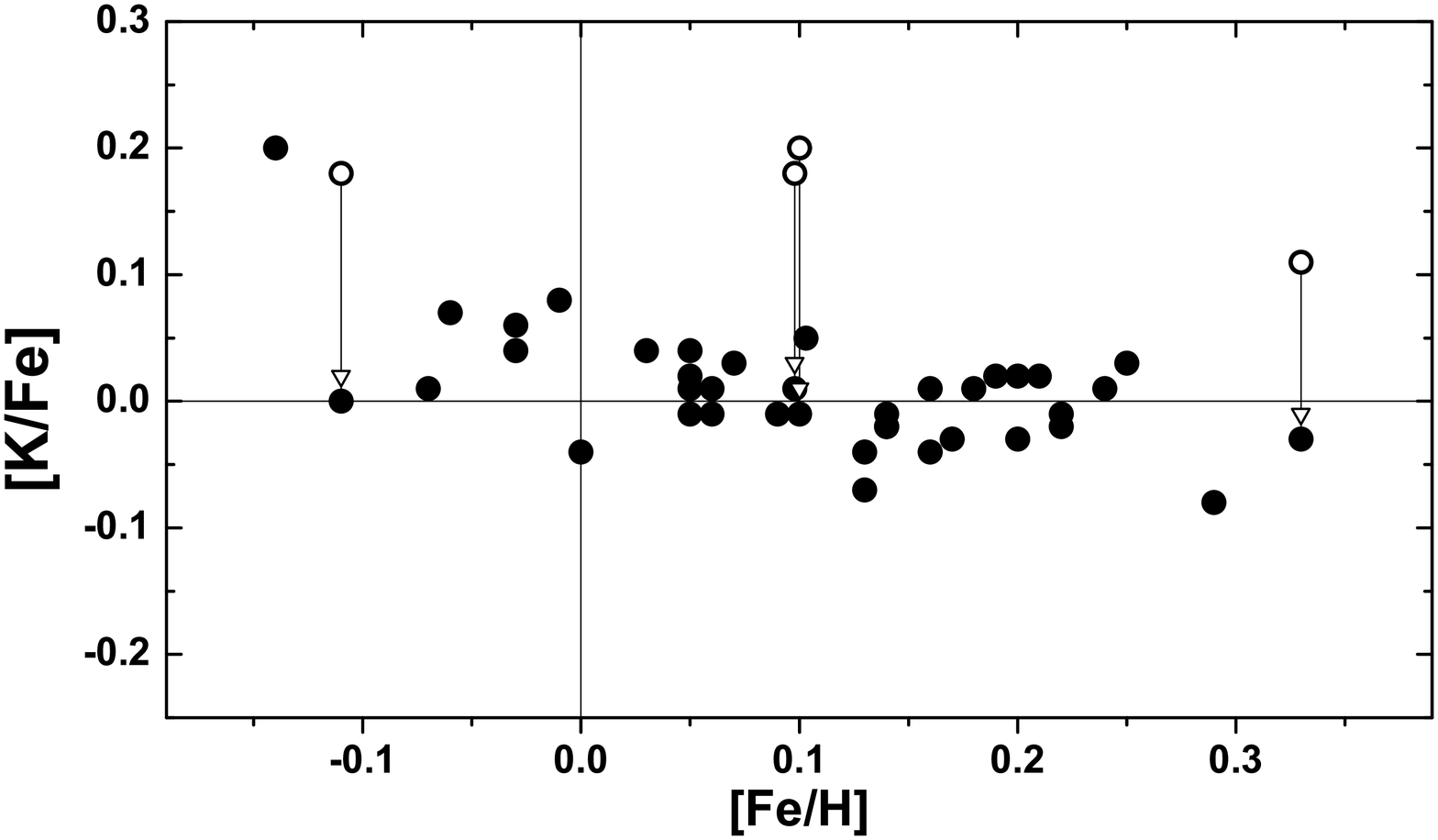}\\
\includegraphics[width=0.98\columnwidth,clip=true]{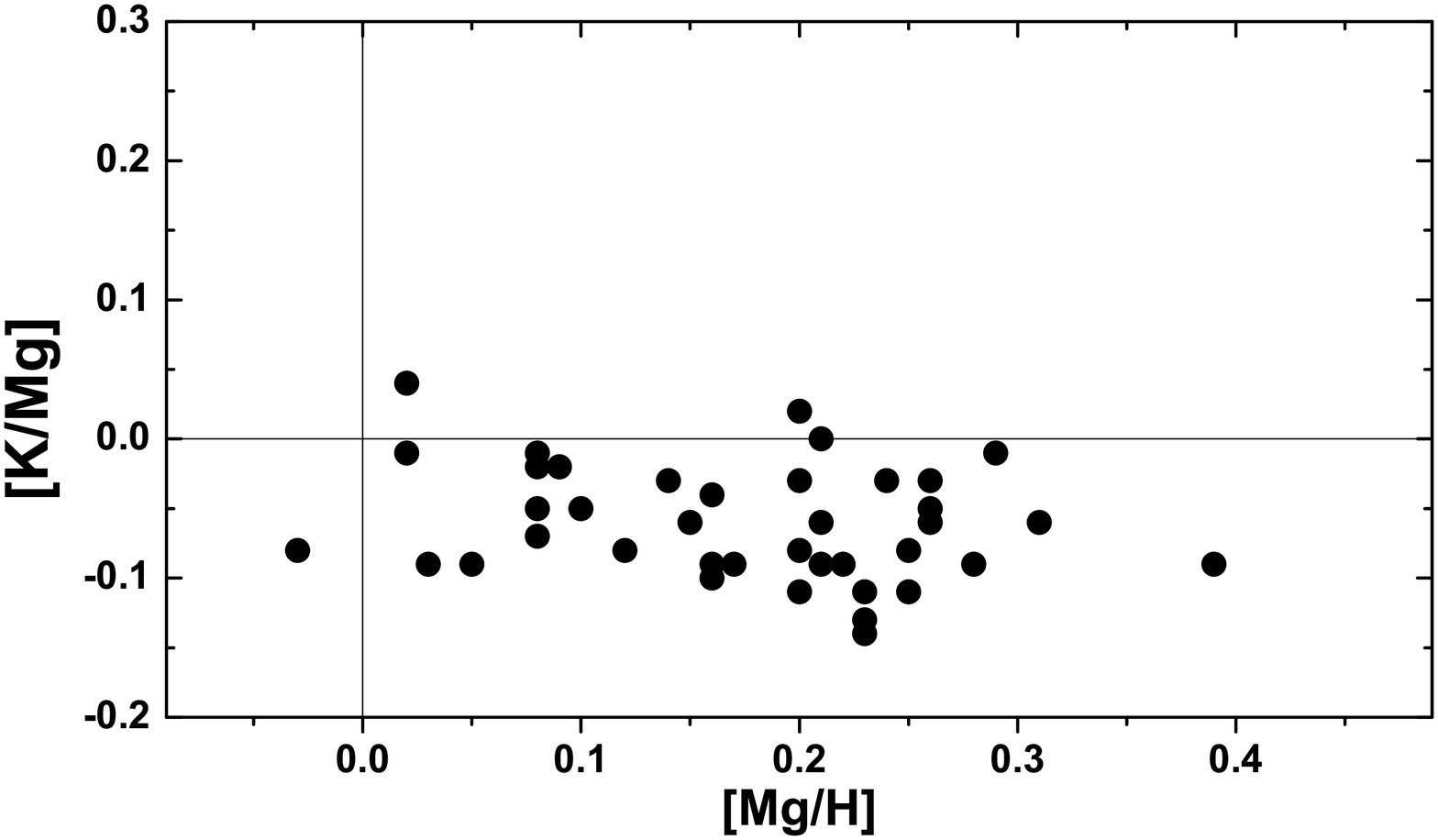}
\caption{Same as Fig~\ref{na_fe} but for potassium.}
\label{k_fe}
\end{figure}

The NLTE correction, we derived for the four selected stars, is not negligible;
it is comparable to or slightly larger than the uncertainty on the abundance determination.
In Appendix (Fig. \ref{linefit1}) we show NLTE and LTE profiles of two potassium lines in the 
solar and stellar spectra. Fig. \ref{k_fe} is the same as Fig. \ref{na_fe} but for potassium. 

\subsection{{Sr}\,{\sc II}} 
The strontium NLTE atomic model is from \citet{Andrievsky11}. Our solar strontium abundance
derived from the solar optical spectrum is A(Sr)=2.92 (see Table\,\ref{sun}). 
Synthetic spectra of the three lines at 1003.6, 1032.7, 1091.4\,nm (\loggf are from \citealt{Warner68} 
and $\Gamma_{vw}$ are from \citealt{Barklem00}) are able to well reproduce the solar observed spectrum. 
We found that NLTE effects strongly influence these lines. As a rule, the line EWs in the solar spectrum are increased by 20--25\%, but in some cases the NLTE corrections achieve about 0.20--0.36\,dex. The mean 
NLTE abundance differs from the LTE abundance published by \citet{Caffau19a} by about 
0.5\,dex. Surely, in this difference the NLTE plays a big role. 
The [Sr/Fe] ratio shows a clear dependence on [Fe/H]: the lower the metallicity, the larger 
the [Sr/Fe] ratio. When [Fe/H]=+0.10, [Sr/Fe] equals to zero. 

\begin{figure}
\includegraphics[width=0.98\columnwidth,clip=true]{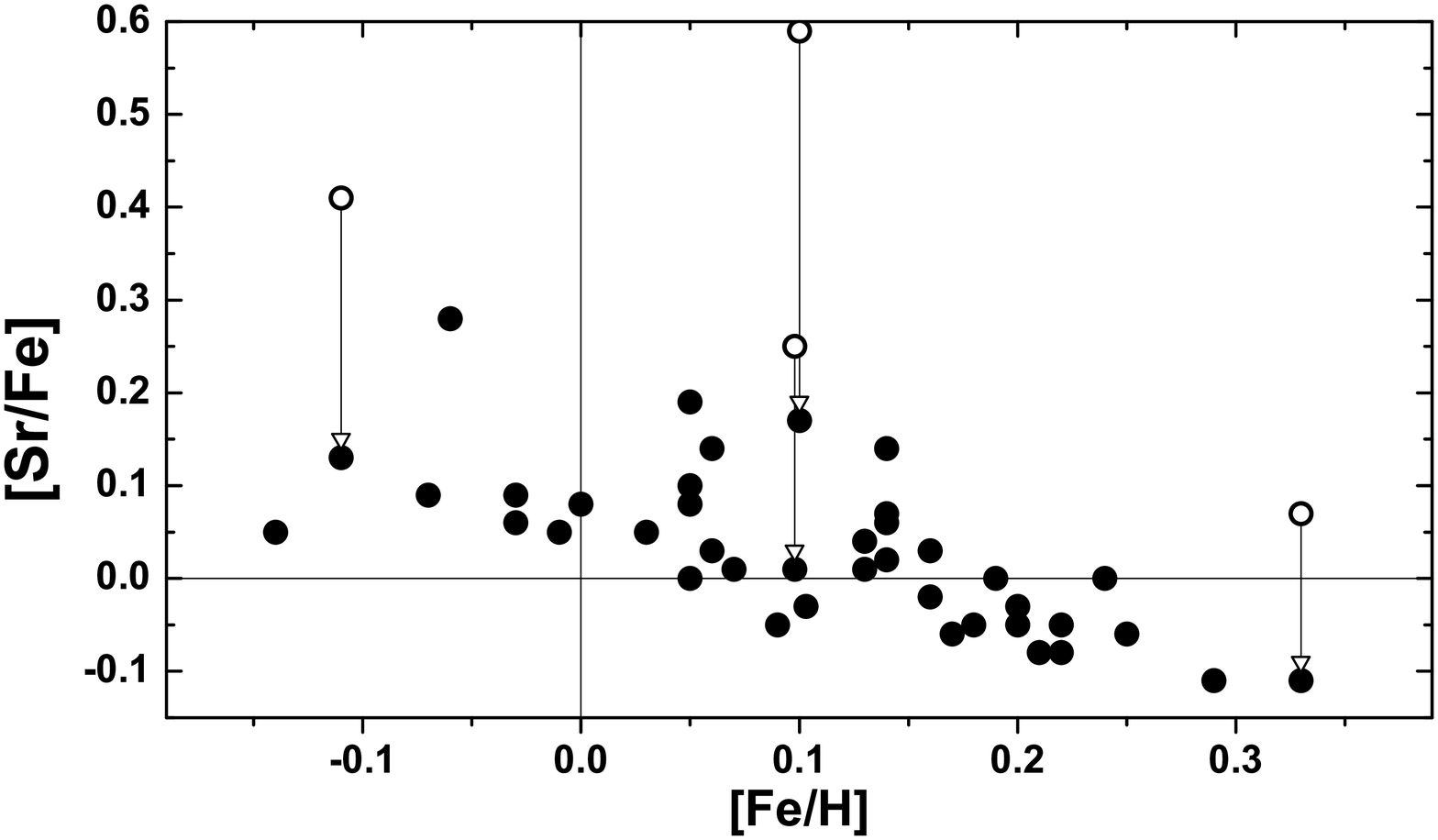}
\includegraphics[width=0.98\columnwidth,clip=true]{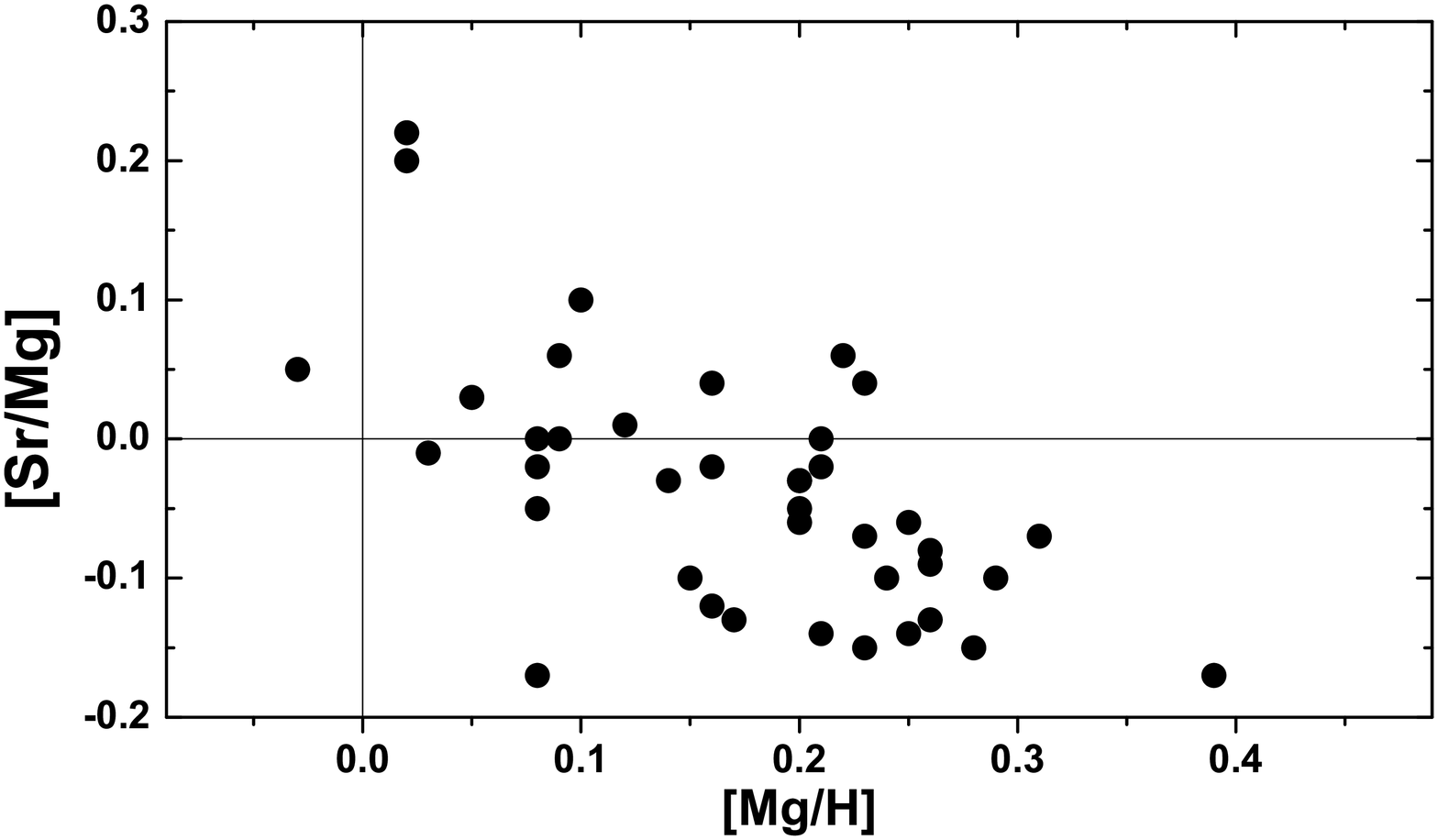}
\caption{Same as Fig.~\ref{na_fe} but for strontium.}
\label{sr_fe}
\end{figure}

The NLTE correction, we derived for the four selected stars, is in the range from --0.18 (for the coolest star)
to --0.45 (for the hottest star). Being the uncertainty on the abundance determination small 
(always within 0.1\,dex), for these Sr lines the NLTE effect should be taken into account, when possible.
In Appendix (Fig. \ref{linefit1}) we show NLTE and LTE profiles of two strontium lines in the solar and stellar 
spectra. Fig.~\ref{sr_fe} is the same as Fig.~\ref{na_fe} but for strontium. 

\section{Discussion}

The main goal of this investigation was to provide, for \ion{Na}{I}, \ion{Al}{I}, 
\ion{Mg}{I}, \ion{S}{I}, \ion{K}{I} and \ion{Sr}{II}, a list of lines in the IR 
wavelength range that can be safely used in an LTE investigation.
The selected lines, covering the complete wavelength range of GIANO (950--2450\,nm), are listed in Table\,\ref{lin}.
In the Table we provide the oscillator strength, the $\Gamma_{vw}$ and a comment on the line:
LTE means the line can safely be analysed in the LTE approximation (absolute NLTE correction is smaller 
than 0.02 dex), while NLTE means that the line is affected by departure from LTE. For a few lines the string ``small NLTE'' means that there is a minor NLTE effect (not larger than 5\% in EW).
These computations can safely be applied for unevolved, solar-metallicity stars, with stellar parameters
similar to the sample here analysed (see Fig\,\ref{fig:hr}). Clearly, in another parameter space NLTE effects can be different.
In Fig.\,\ref{linefit1} for a few selected lines the observed profiles of the Sun and of the
three stars (HD\,20670, HD\,24040 and HD\,114174) are compared to the LTE and NLTE theoretical profiles.
As one can see, the agreement of NLTE synthetic profile with the observed spectrum is very good, while for
several cases the theoretical LTE profile deviates from the observation.

The trends of the investigated elements as a function of the metallicity 
confirm the ones that are apparent from the APOGEE data, although 
there are some minor differences in the slopes for Na, Mg and S.
\begin{itemize}
\item{Na.} In the upper panel of Fig.\,\ref{na_fe}, [Na/Fe] is plotted as a function of [Fe/H], using Fe as a proxy for metallicity.
The four stars with an LTE analysis (open symbols in the figure, connected with a line to the NLTE result) show the small NLTE effects for this parameters space. In fact, of the twelve \ion{Na}{i} lines investigated, only three are sensitive to NLTE (see Table\,\ref{lin}). In the lower panel of the same figure, Mg is used as a proxy for the metallicity, taking the advantage of a more coherent picture because both Na and Mg are investigated in NLTE.
In both panels, a relative increase in the ratio of Na over metallicity at increased metallicity is visible.
In Fig.\,\ref{nafecomp} the analysis here done is compared to the analysis by \citet{Caffau19a} on the same stars.
The average difference from the two investigations is 0.09\,dex, ranging from about zero to 0.25\,dex.
This difference is not due to NLTE effects (which is of the order of 0.03\,dex), but to the
adjustment of the atomic data on the solar spectrum done here and not in \citet{Caffau19a}.
Surely with this new analysis the scatter star-to-star is smaller.
Summarizing, a great majority of sodium lines in the wide range of 
effective temperature from 5000 to 6500\,K and sodium relative abundances 
[Na/H] from --0.8 to +0.8 are insensitive to the deviation from LTE. This is 
valid for the dwarfs, but for the stars with \logg\ less than 3 one needs to 
apply the deatiled NLTE calculatins.

\begin{figure}
\includegraphics[width=0.98\columnwidth,clip=true]{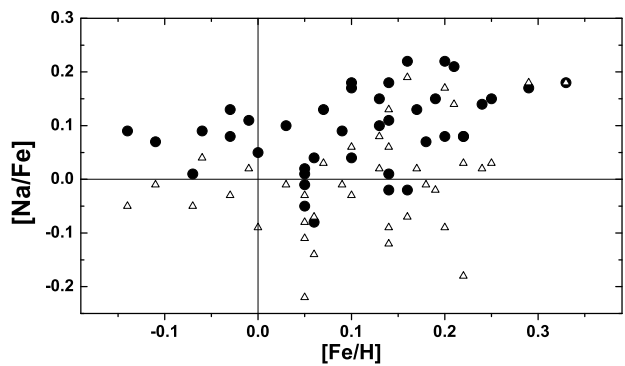}
\caption{Relative sodium abundances ([Na/Fe] versus [Fe/H] from this analysis (filled circles)
compared to the results of analysis \citet{Caffau19a} (open tiangles).}
\label{nafecomp}
\end{figure}
\item{Mg.} As visible in Fig.\,\ref{mg_fe} and in Table\,\ref{dNLTE}, the NLTE 
effects for Mg are tiny. Of the 26 \ion{Mg}{i} lines (the eight components of 1081.1\,nm we 
considered as a single line) investigated, four are sensitive to NLTE (see  Table\,\ref{lin}).
As visible from the Figure, there is an evident trend of the [Mg/Fe] to decreasing at increasing metallicity.
In general, in the effective temperature range from 5200 to 6500\,~K the 
NLTE effects are the same as previously described (for relative magnesium 
abundance [Mg/H] from --0.7 to +o.7). However, the deviations increase for the 
giant stars with \logg\ less than 2.5.

\item{Al.} From the upper panel of Fig.\,\ref{al_fe}, one can see a flat [Al/Fe] versus [Fe/H], while on the lower
panel, there is a hint for an incresing [Al/Fe] for increasing [Mg/Fe].
The NLTE effects on the \ion{Al}{i} lines are about 0.05\,dex (see Table\,\ref{dNLTE}), smaller than the line-to-line scatter for all stars. The nine \ion{Al}{i} lines here analyzed suffer all but one from NLTE effects, 
which are anyway not strong. The mean difference from the analysis of \citet{Caffau19a} is 0.05\,dex, with 
a few cases with differences up to 0.15\,dex that are due to atomic data and line selection.
As a concluding remark it can be noted that two \ion{Al}{i} lines 1087.2 
and 1089.1\,nm can be safely used in LTE analysis for the dwarfs stars with 
\logg\ larger than 3 and affective temperature lower than 6000\,K
(for the wide range of the relative aluminum abundances from --0.7 to +0.7). 
At the same time for the subgiant and giant stars, as well as for hotter dwarfs 
the detailed NLTE calculations are necessary.

\item{S.} In the upper panel of  Fig.\,\ref{s_fe}, the decrease of [S/Fe] with increasing metallicity typical 
in the $\alpha$-elements, is well evident. In the lower panel of Fig.\,\ref{s_fe}, an almost flat [S/Mg] as a function
of [Mg/Fe] as a proxy for metallicity is  expected because both Mg and S are $\alpha$-elements.
The behaviour was very similar in \citet{Caffau19a}, also if single stars could differ up to 0.17\,dex, due to choice
of the lines, atomic data, but also NLTE for the lines belonging to the multiplet\,3.
Nineteen lines of \ion{S}{i} have been analyzed (see Table\,\ref{lin}). Only the three lines belonging to the multiplet\,3 (the three bluest lines here analyzed) are strongly affected by NLTE; the other lines are either well represented in LTE or slightly affected by NLTE. The lines of multiplet\,3 are the strongest and sometimes the only possibility to derive A(S) in metal-poor stars.
The global NLTE effects on the abundance determination for S is small (up to --0.08, see Table\,\ref{dNLTE})
because  all the \ion{S}{i} lines, but the three of multiplet\,3, form in conditions close to LTE.
Speaking of  sensitivity to  atmosphere parameters, one can note that 
seven \ion{S}{i} lines at 2250.7--2270.7\,nm show the weak influence from the 
NLTE effects if we are considering the dwarfs with \logg\ greater than 3 and 
effective temperatures lower than 6400\,K (relative sulfur abundance is 
from --0.7 to +0.7). NLTE effects are becoming remarkable for the subgiant and 
giant stars. The same is true for the metal-deficient stars ([Fe/H] $<$ --0.7).

\item{K.} From both panels of Fig.\,\ref{k_fe}, one can see a decreasing [K/Fe] for
increasing metallicity.
Of the seven \ion{K}{i} lines here analyzed, four are affected by NLTE (see Table\,\ref{lin}).
The NLTE correction for the stars listed in Table\,\ref{dNLTE} is larger than the line-to-line scatter and the
large difference of the LTE and NLTE abundance is evident in Fig.\,\ref{k_fe}.
The difference with the analysis by \citet{Caffau19a} is large, certainly affected by the NLTE.
Our recommendation for the use of potassium lines depending on the parameters of 
stellar atmospheres is as follows: lines at 1516.3--1516.8\,nm are available 
for use in the LTE approximation for stars with \logg $>$ 3.5 and [K/H] from 
--0.7 to +0.7. The effective temperature has almost no effect on the NLTE 
corrections for these lines.

\item{Sr.} Fig.\,\ref{sr_fe} shows a decrease of [Sr/Fe] or [Sr/Mg] for increasing metallicity.
The three lines here analyzed are strongly affected by NLTE (see Table\,\ref{lin}) and the NLTE corrections
for the stars with both LTE and NLTE analysis are much larger than the line-to-line scatter 
and can be up to 0.45\,dex (see Table\,\ref{dNLTE}).

Concluding our discussion, we should make the following remark: we recommend using 
certain infrared lines to obtain elemental abundances in LTE, while at the same time leaving 
open questions about the effect of the limitations of our approach, such as the 1-D approximation, 
on the obtained abundances.

\end{itemize}

\section{Conclusion}

We analyzed IR lines of several ions in the solar spectrum and in the spectra of 
40 dwarf stars observed with GIANO spectrograph. We have studied in detail the
influence of the NLTE effects on the lines under consideration, and we give 
our recommendation on which lines can be used to obtain abundance in the LTE 
approximation, and which lines require NLTE processing. This information is 
given in the Table \ref{lin} in the Appendix.

The sample here discussed is small in size, but has a high quality
of the spectra, both in terms of S/N ratio and resolution.
It is interesting to note how such a small size sample 
allows to derive reliable abundance trends that become apparent only
for samples that are several orders of magnitude larger in size,
but of lower quality, like the APOGEE sample.

\section*{Acknowledgements}
We would like to thank the referee of this paper for his/her valuable comments.



\appendix

\section{}

\begin{figure*}
  \centering
  \begin{minipage}[b]{0.45\textwidth}
    \includegraphics[width=\textwidth]{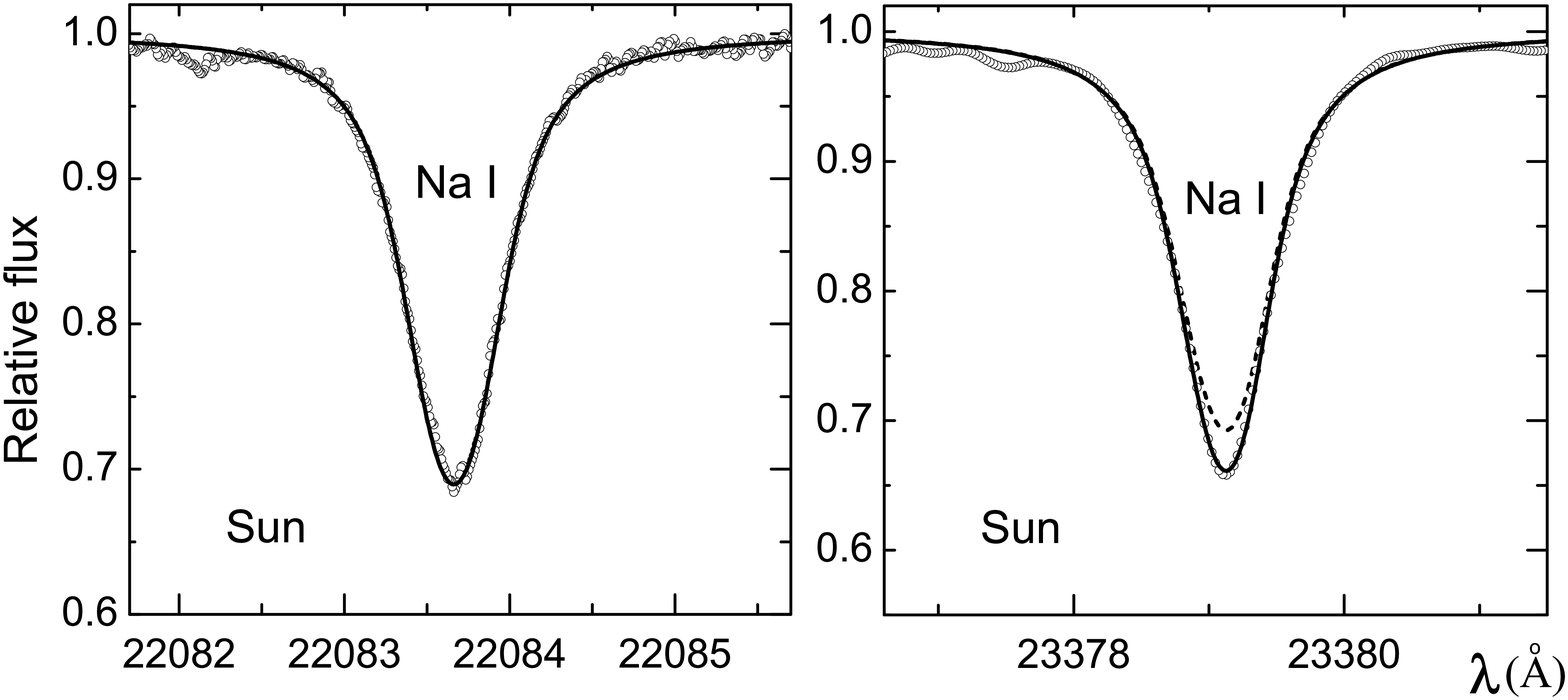}
  \end{minipage}
  \begin{minipage}[b]{0.45\textwidth}
    \includegraphics[width=\textwidth]{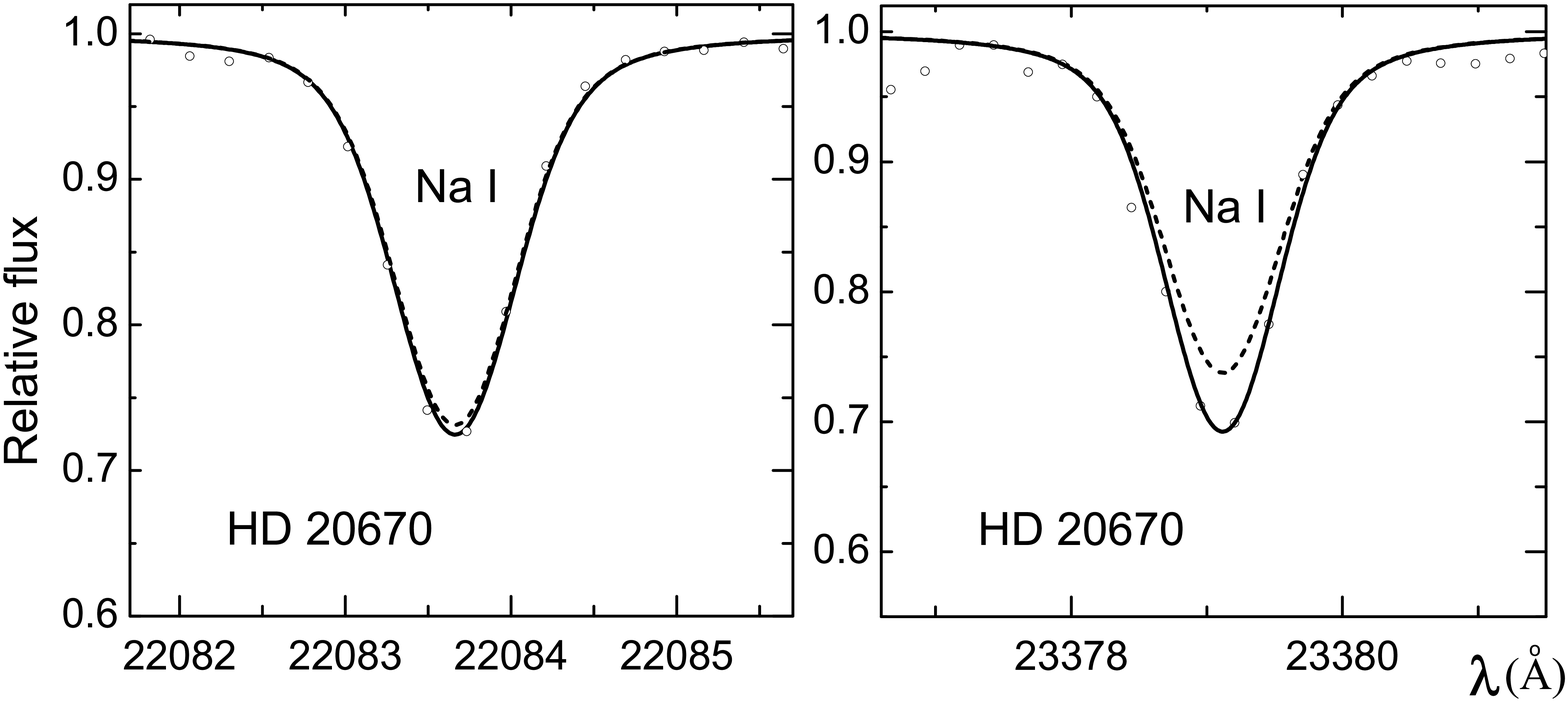}
  \end{minipage}
  \begin{minipage}[b]{0.45\textwidth}
    \includegraphics[width=\textwidth]{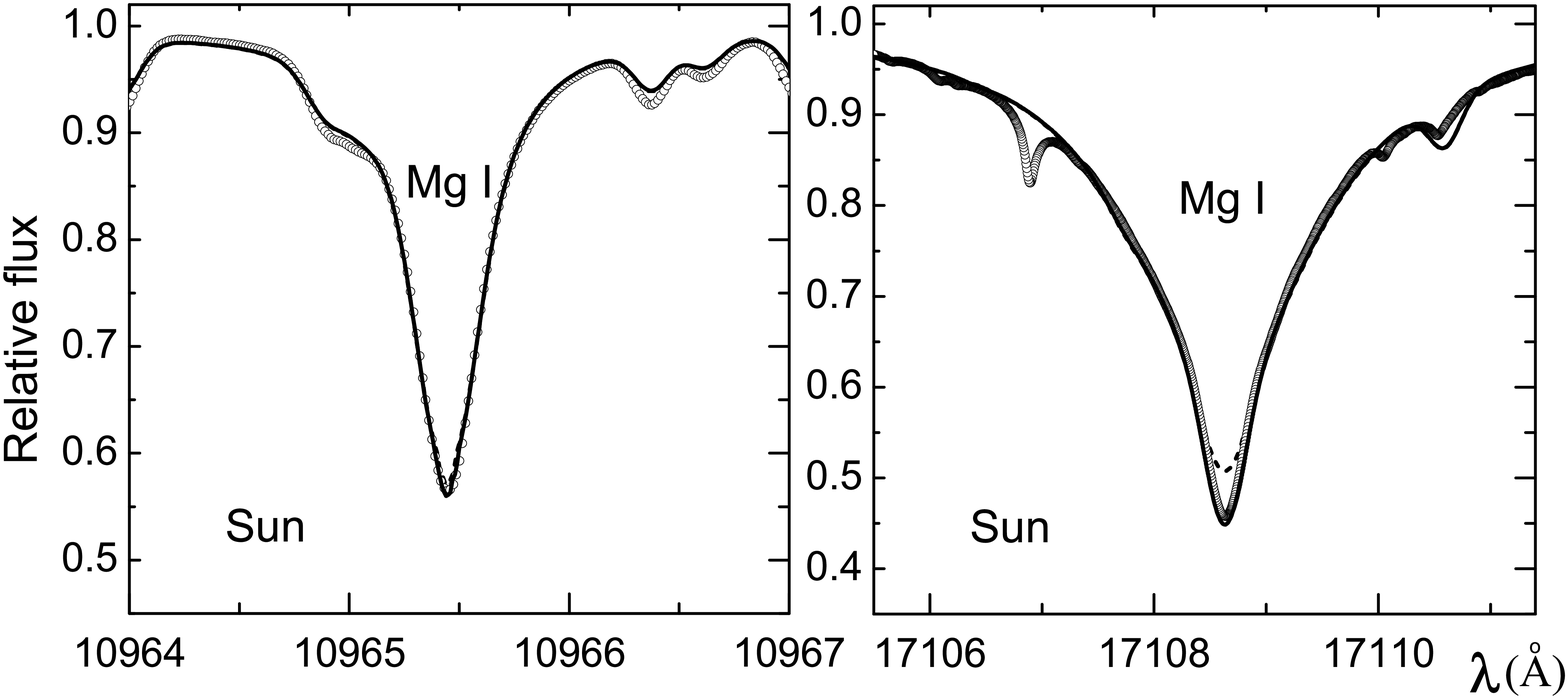}
  \end{minipage}
  \begin{minipage}[b]{0.45\textwidth}
    \includegraphics[width=\textwidth]{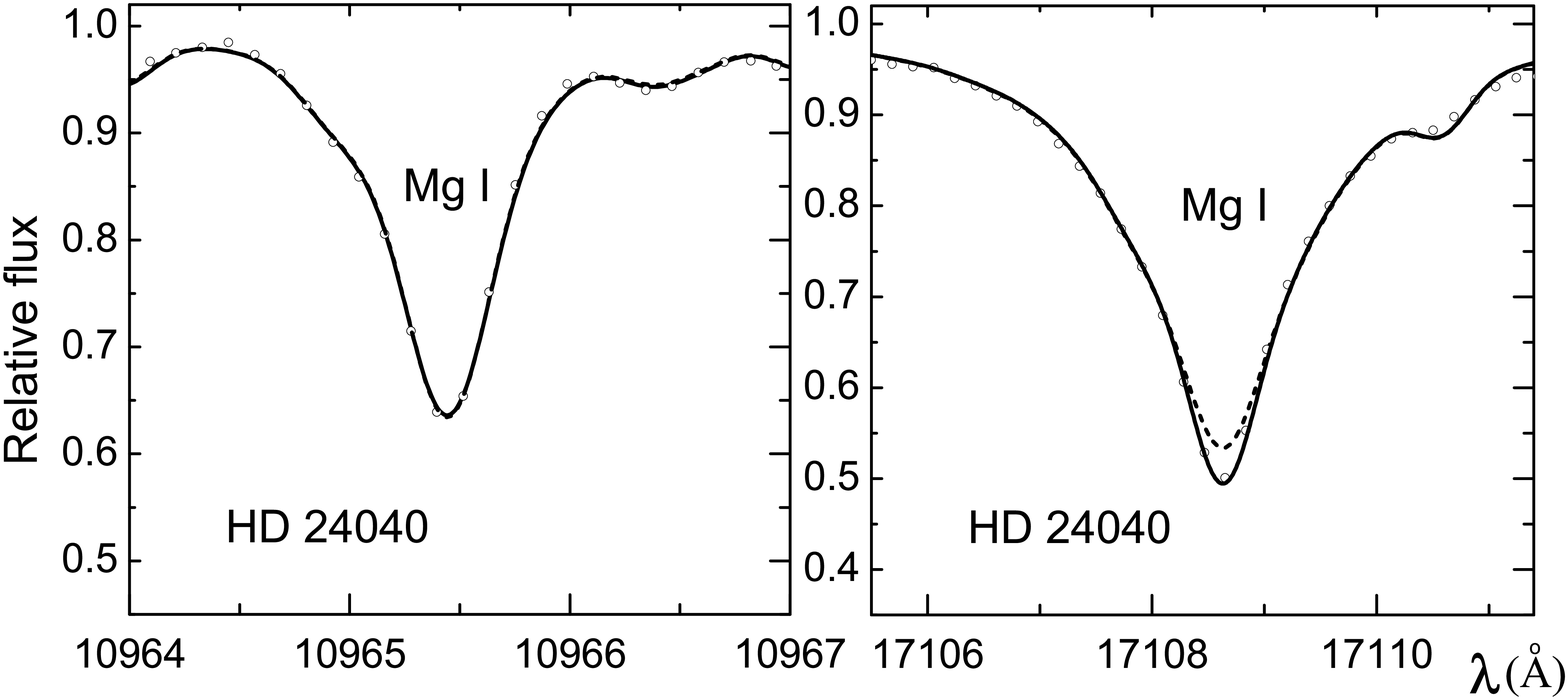}
  \end{minipage}
  \begin{minipage}[b]{0.45\textwidth}
    \includegraphics[width=\textwidth]{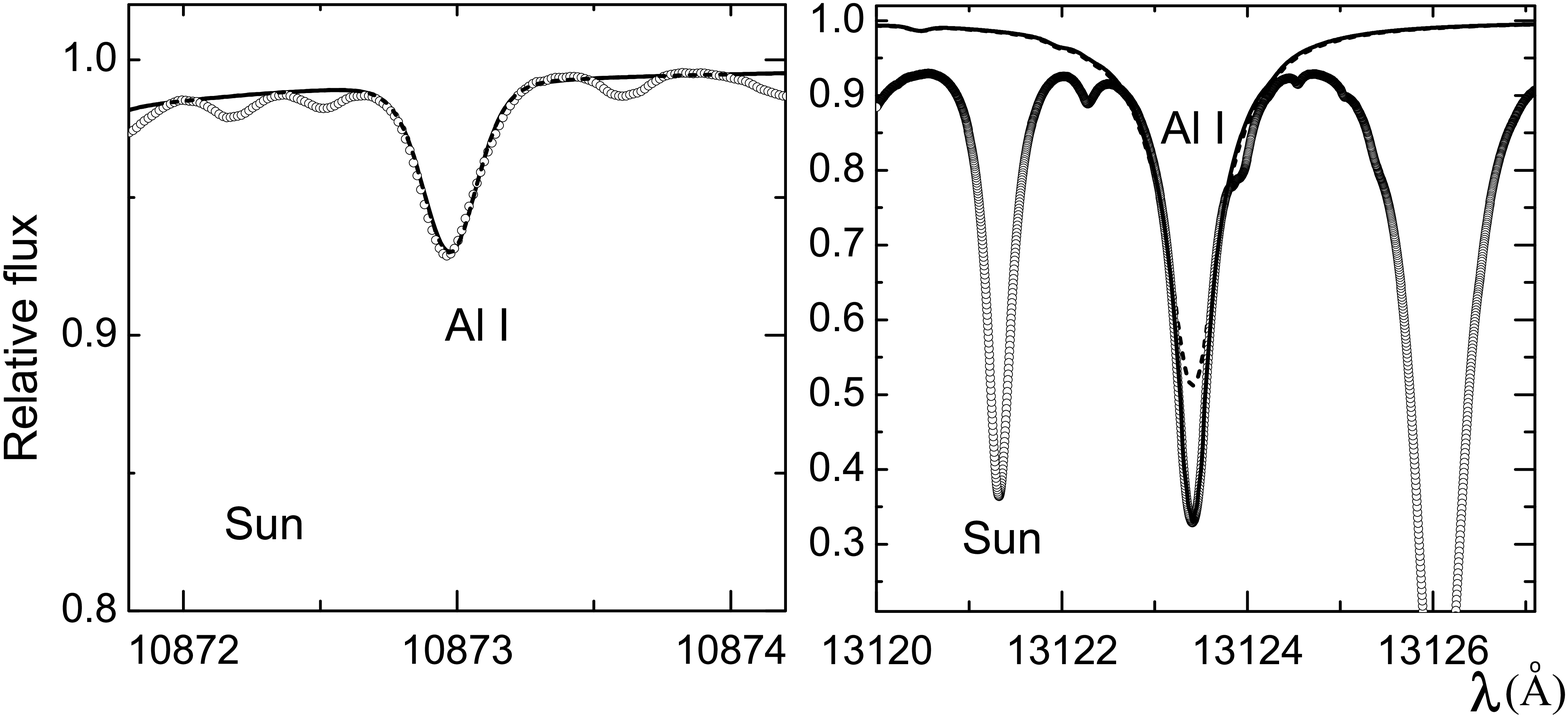}
  \end{minipage}
  \begin{minipage}[b]{0.45\textwidth}
    \includegraphics[width=\textwidth]{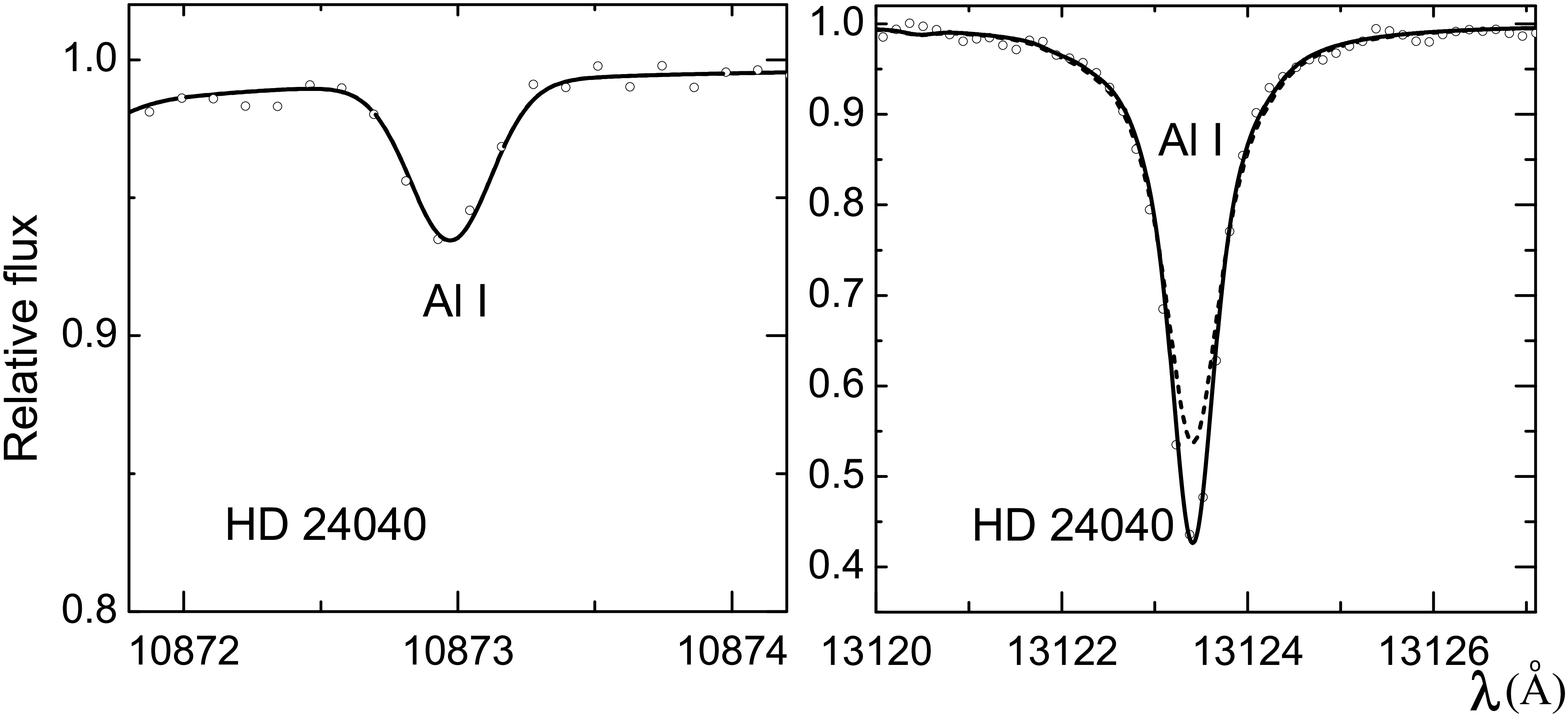}
  \end{minipage}
  \begin{minipage}[b]{0.45\textwidth}
    \includegraphics[width=\textwidth]{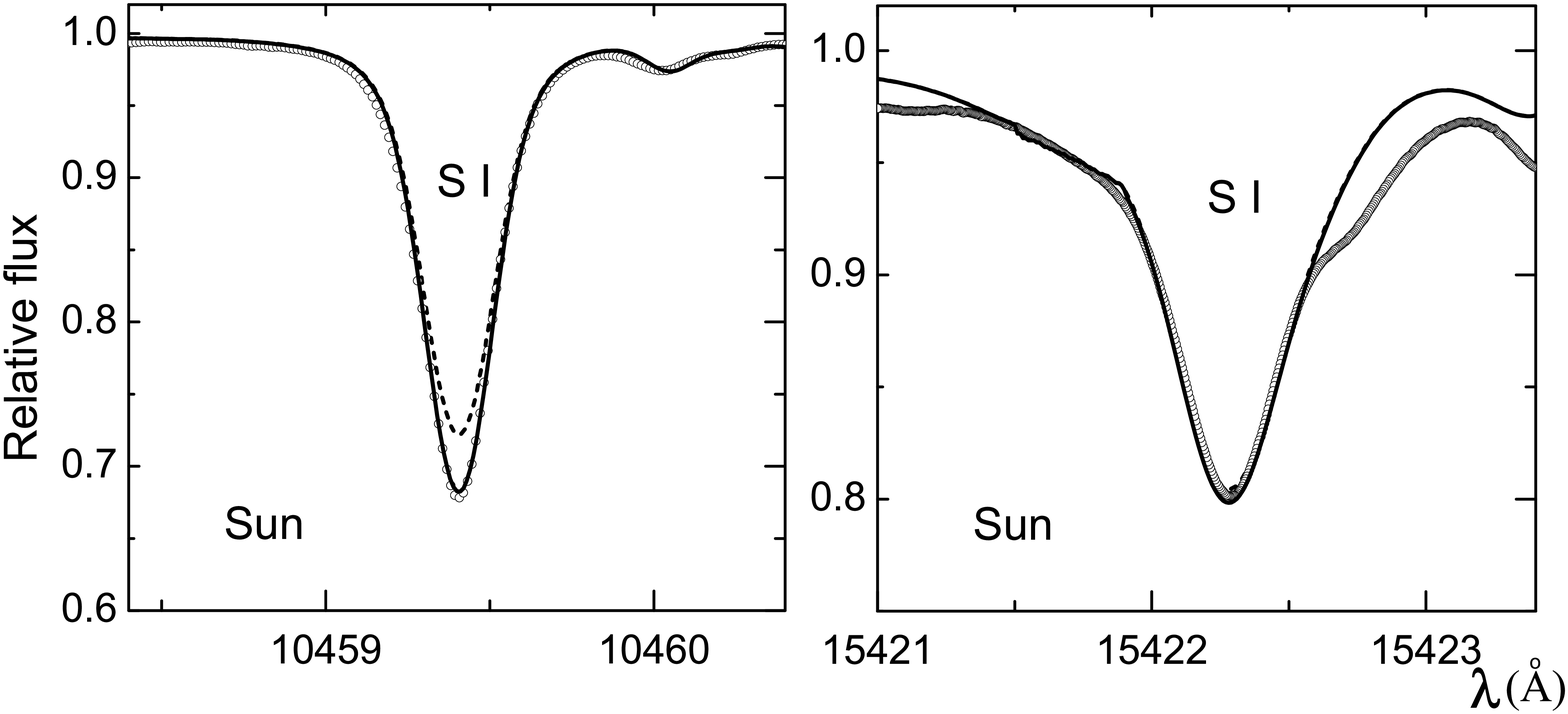}
  \end{minipage}
  \begin{minipage}[b]{0.45\textwidth}
    \includegraphics[width=\textwidth]{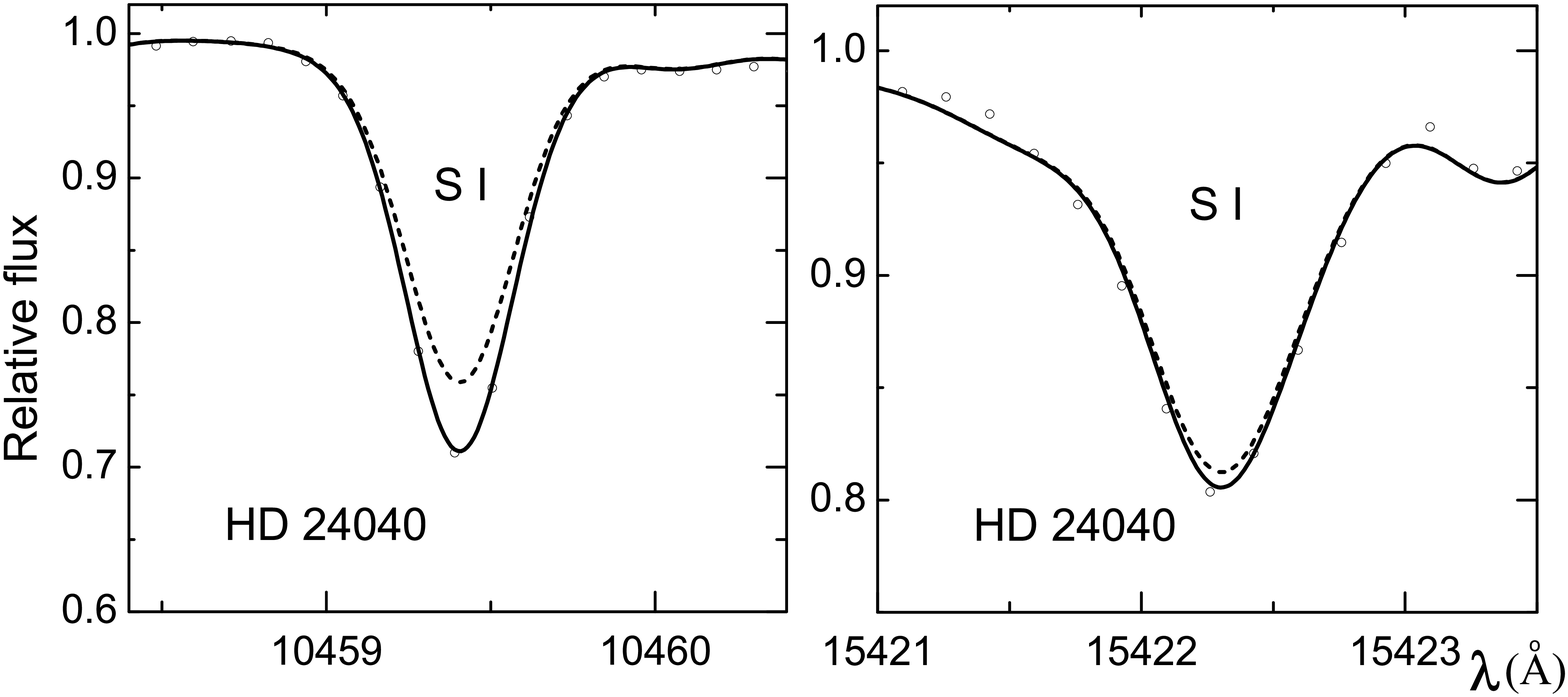}
  \end{minipage}
  \begin{minipage}[b]{0.45\textwidth}
    \includegraphics[width=\textwidth]{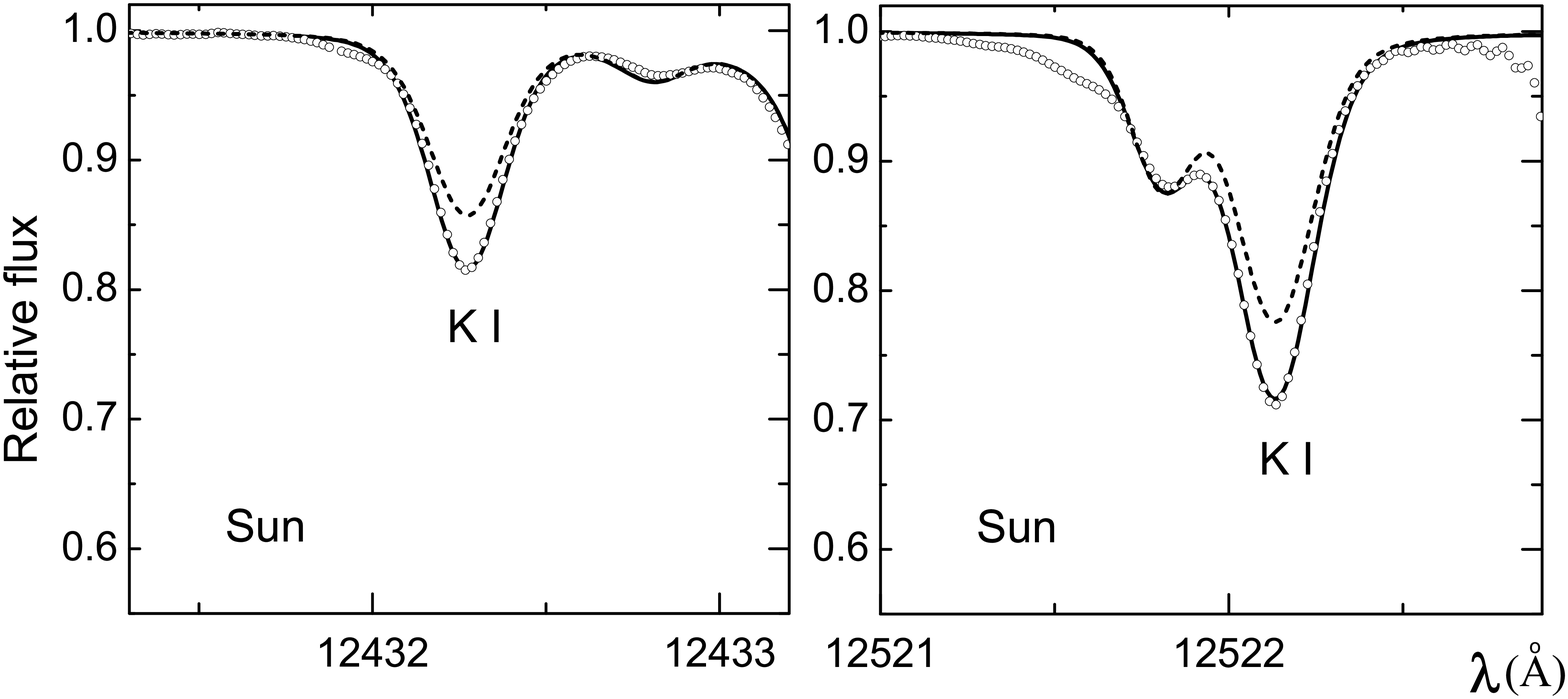}
  \end{minipage}
  \begin{minipage}[b]{0.45\textwidth}
    \includegraphics[width=\textwidth]{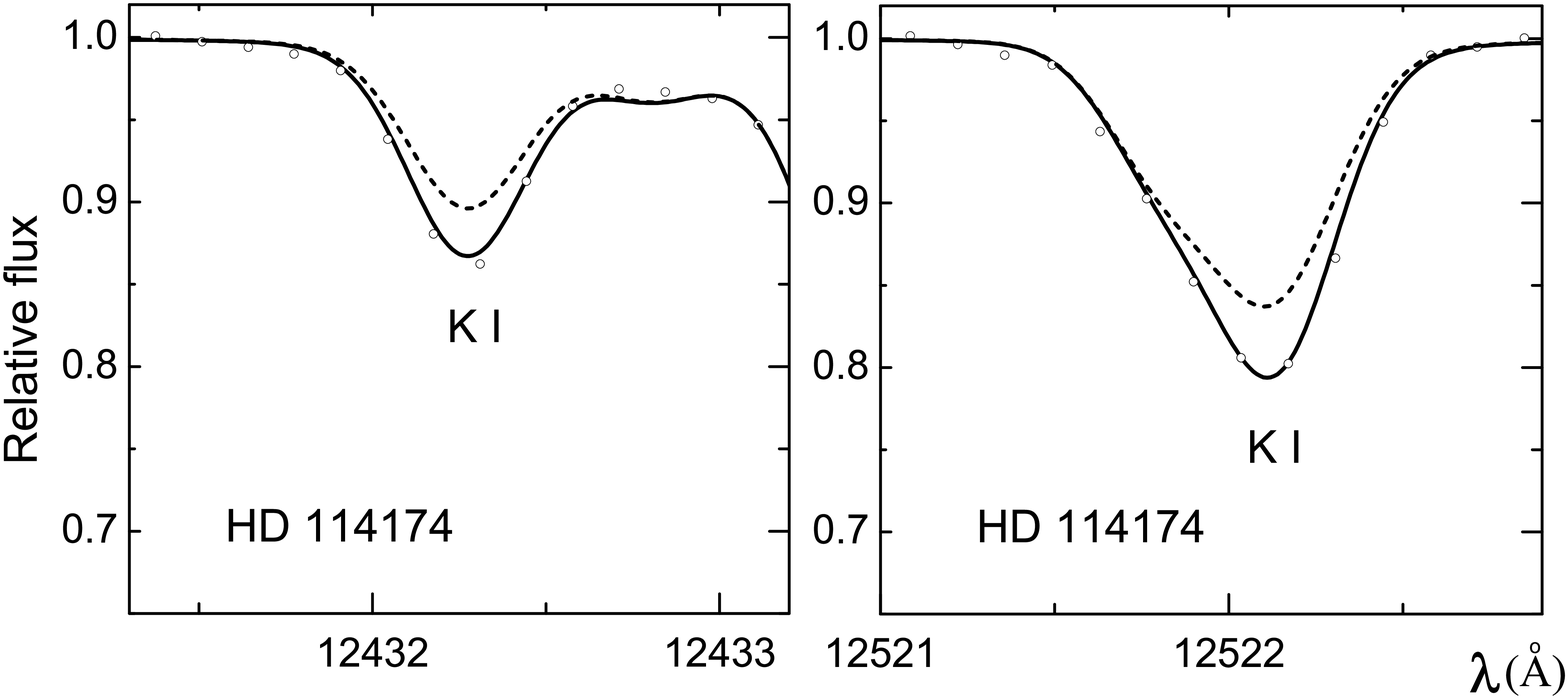}
  \end{minipage}
  \begin{minipage}[b]{0.45\textwidth}
    \includegraphics[width=\textwidth]{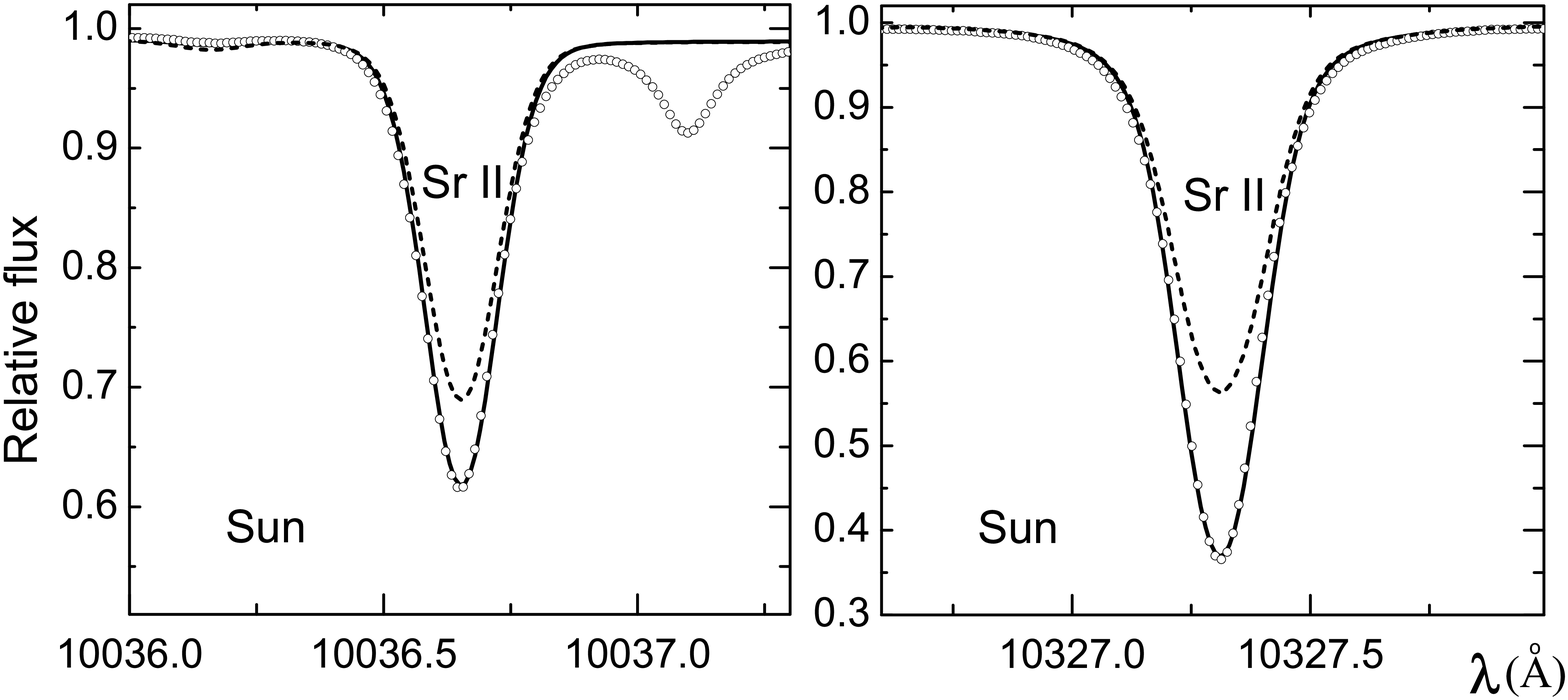}
  \end{minipage}
  \begin{minipage}[b]{0.45\textwidth}
    \includegraphics[width=\textwidth]{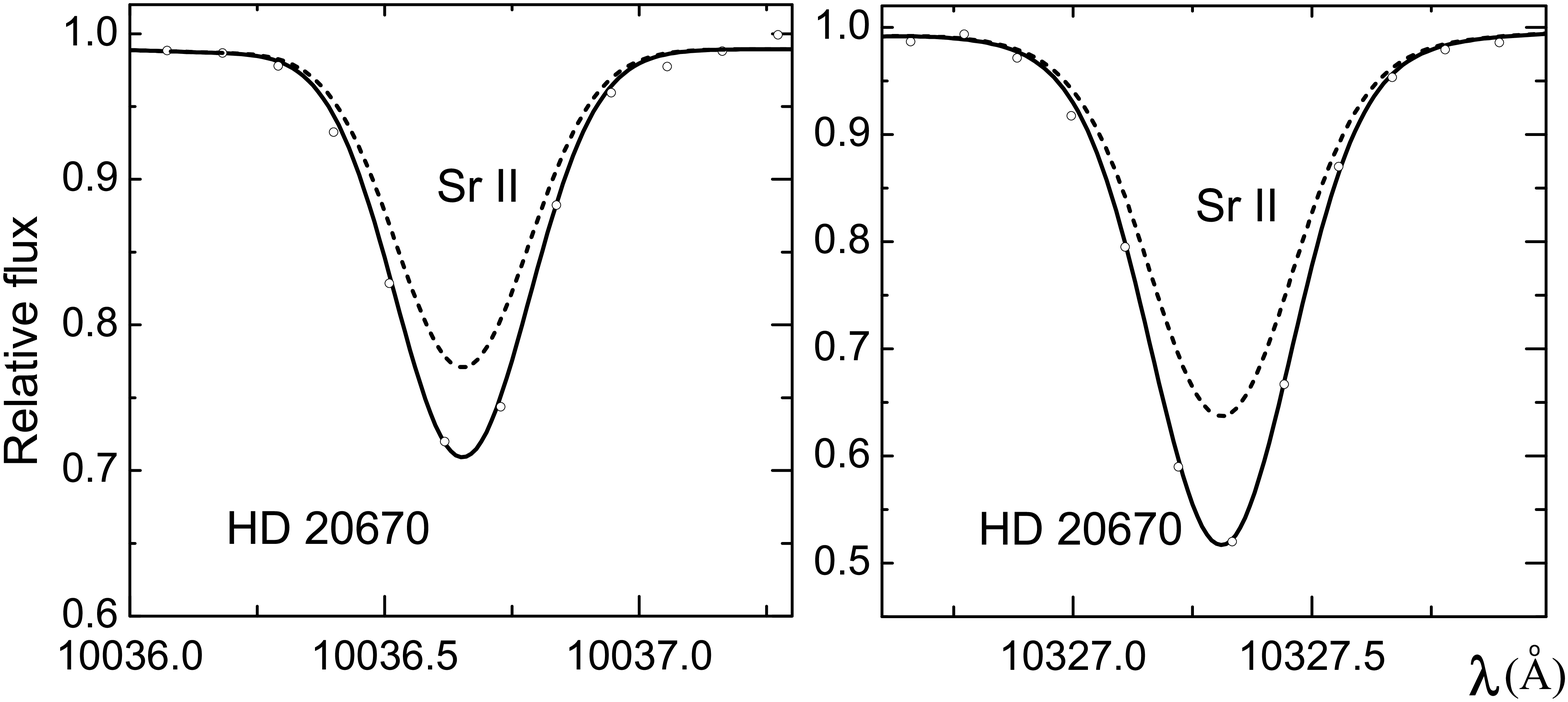}
  \end{minipage}
\caption{Observed fragments of the spectra ($open~circles$) and synthesized profiles of two selected \ion{Na}{i}, \ion{Mg}{i}, \ion{Al}{i}, \ion{S}{i}, 
\ion{K}{i}, \ion{Sr}{ii} lines. LTE ($dashed~line$) and NLTE ($continuous~line$).}
\label{linefit1}
\end{figure*}

\begin{table*}
\caption{Studied lines and our recommendation: LTE or NLTE.}
\label{lin}
\begin{tabular}{rrcl|rrcl}
\hline
$\lambda$,\AA&\loggf&$\Gamma_{vw}$&LTE/NLTE&$\lambda$,\AA&\loggf&$\Gamma_{vw}$&LTE/NLTE\\
\hline
\hline
&&&&&&&\\
Na I	 &  	  &        &            &Al I	 &  	  &        &             \\
10746.45 &  --1.29 &  --7.06 &  LTE       &10872.97 &  --1.33 &  --7.19 &  LTE       \\
10834.85 &  --0.50 &  --7.06 &  LTE       &10891.74 &  --1.03 &  --7.19 &  LTE       \\
10834.85 &  --1.80 &  --7.06 &  LTE       &13123.41 &   0.27 &  --7.10 &  NLTE      \\
10834.91 &  --0.66 &  --7.06 &  LTE       &13150.75 &  --0.03 &  --7.10 &  NLTE      \\
12679.17 & --0.04 &  --6.65 &  LTE       &16718.96 &   0.22  &  --7.16 &  NLTE      \\
12679.17 &  --1.34 &  --6.65 &  LTE       &16750.56 &   0.47 &  --7.16 &  NLTE      \\
12679.22 &  --0.20 &  --6.65 &  LTE       &16763.36 &  --0.48 &  --7.16 &  small NLTE\\
22056.40 &   0.29 &  --7.06 &  LTE       &21093.03 &  --0.31  &  --6.90 &  NLTE      \\
22083.66 &  --0.01 &  --7.06 &  LTE       &21163.76 &  --0.01 &  --6.90 &  NLTE      \\
23348.38 &   0.28 &  --7.06 &  NLTE      &         &        &        &            \\
23378.96 &  --0.42 &  --7.06 &  NLTE      &S I	  &  	  &        &             \\
23379.14 &   0.54 &  --7.06 &  NLTE      &10455.45 &   0.25 &  --7.32 &  NLTE      \\
         &        &        &             &10456.76 & --0.45 &  --7.32 &  NLTE      \\
Mg I	 &  	  &        &             &10459.41 &   0.03 &  --7.32 &  NLTE      \\
 9983.19 &  --2.15 &  --7.01 &  LTE      &15400.08 &   0.45 &  --7.31 &  small NLTE\\
 9986.48 &  --1.68 &  --7.01 &  LTE      &15403.72 & --0.28 &  --7.31 &  small NLTE\\
 9993.21 &  --1.45 &  --7.01 &  LTE      &15403.79 &   0.63 &  --7.31 &  small NLTE\\
10811.05 &   0.02 &  --6.82 &  NLTE      &15422.20 & --1.82 &  --7.31 &  small NLTE\\
10811.08 &  --0.14 &  --6.82 &  NLTE     &15422.26 & --0.28 &  --7.31 &  small NLTE\\
10811.10 &  --1.04 &  --6.82 &  NLTE     &15422.28 &   0.79 &  --7.31 &  small NLTE\\
10811.12 &  --1.04 &  --6.82 &  NLTE     &15469.82 & --0.15 &  --7.40 &  small NLTE\\
10811.14 &  --2.59 & --6.82 &  NLTE      &15475.62 & --0.62 &  --7.40 &  small NLTE\\
10811.16 &  --0.30 &  --6.82 &  NLTE     &15478.48 &   0.08 &  --7.40 &  small NLTE\\
10811.20 &  --0.19 &  --6.82 &  NLTE      &22507.56 &  --0.48 &  --7.61 &  LTE       \\
10811.22 &  --1.28 &  --6.82 &  NLTE      &22519.07 &  --0.25 &  --7.61 &  LTE       \\
10953.32 &  --0.86 &  --6.78 &  LTE       &22552.57 &  --0.04 &  --7.61 &  LTE       \\
10957.28 &  --0.99 &  --6.78 &  LTE       &22563.83 &  --0.26 &  --7.61 &  LTE       \\
10957.30 &  --0.51 &  --6.78 &  LTE       &22575.39 &  --0.73 &  --7.61 &  LTE       \\
10965.39 &  --2.16 &  --6.78 &  LTE       &22644.06 &  --0.34 &  --7.61 &  LTE       \\
10965.41 &  --0.99 &  --6.78 &  LTE       &22707.74 &   0.44 &  --7.61 &  LTE       \\
10965.45 &  --0.24 &  --6.78 &  LTE       &         &       &         &            \\
11828.17 &  --0.33 &  --7.19 &  NLTE      &K I	 &  	  &        &             \\
12039.82 &  --1.53 &  --7.22 &  LTE       &11769.64 &  --0.48 &  --7.33 &  NLTE      \\
12083.28 &  --0.79 &  --7.12 &  NLTE      &11772.84 &   0.47 &  --7.33 &  NLTE      \\
12083.65 &   0.41 &  --7.12 &  NLTE      &12432.27 &  --0.43 &  --7.02 &  NLTE      \\
12417.94 &  --1.66 &  --7.15 &  LTE       &12522.13 &  --0.13 &  --7.02 &  NLTE      \\
12423.03 &  --1.19 &  --7.15 &  LTE       &15163.07 &   0.63 &  --7.31 &  LTE       \\
12433.45 &  --0.97 &  --7.15 &  LTE       &15163.07 &  --0.67 &  --7.31 &  LTE       \\
15025.00 &   0.36 &  --7.05 &  small NLTE&15168.38 &   0.48 &  --7.31 &  LTE       \\
15040.25 &   0.14 &  --7.05 &  small NLTE&         &        &        &            \\
15047.71 &  --0.34 &  --7.05 &  small NLTE&Sr II	 &  	  &        &             \\
15740.71 &  --0.21 &  --6.95 &  small NLTE&10036.65 &  --1.31 &  --7.63 &  NLTE      \\
15748.89 &  --0.34 &  --6.95 &  small NLTE&10327.31 &  --0.35 &  --7.63 &  NLTE      \\
15748.99 &   0.14 &  --6.95 &  small NLTE&10914.88 &  --0.64 &  --7.63 &  NLTE      \\
15765.65 &  --1.51 &  --6.95 &  small NLTE&&&&\\
15765.75 &  --0.34 &  --6.95 &  small NLTE&&&&\\
15765.84 &   0.41 &  --6.95 &  small NLTE&&&&\\
17108.63 &   0.06 &  --6.98 &  NLTE      &&&&\\
\hline                                                                                            
\end{tabular}                                                                               
\end{table*}


\bsp	
\label{lastpage}
\end{document}